\documentclass[lettersize,journal]{IEEEtran}
\usepackage{amsmath,amsfonts}
\usepackage{algorithm}
\usepackage{enumerate}
\usepackage{algpseudocode}
\usepackage{array}
\usepackage[caption=false,labelformat=parens, subrefformat=parens]{subfig}
\usepackage{textcomp}
\usepackage{stfloats}
\usepackage{url}
\usepackage{diagbox}
\usepackage{verbatim}
\usepackage{graphicx}
\usepackage{cite}
\usepackage{booktabs}
\usepackage[hidelinks]{hyperref} 
\usepackage{makecell}
\usepackage{xcolor}
\usepackage[export]{adjustbox}
\usepackage{array, multirow}
\usepackage{tabularx}
\usepackage[switch]{lineno}
\newcolumntype{?}{!{\vrule width 1pt}}
\newcommand{\multiline}[2]{%
  \begin{tabularx}{#1\textwidth}[t]{@{}X@{}}
    #2
  \end{tabularx}}
\begin{document}
% \linenumbers
\title{radarODE: An ODE-Embedded Deep Learning Model for Contactless ECG Reconstruction from Millimeter-Wave Radar}

\author{Yuanyuan~Zhang, Runwei~Guan, Lingxiao~Li, Rui~Yang, \IEEEmembership{Senior~Member,~IEEE}, \\ Yutao~Yue, \IEEEmembership{Senior~Member,~IEEE}, Eng~Gee~Lim, \IEEEmembership{Senior~Member,~IEEE}% <-this % stops a space
\thanks{This work has been approved by University Ethics Committee, and is partially supported by Suzhou Science and Technology Programme (SYG202106), Jiangsu Industrial Technology Research Institute (JITRI) and Wuxi National Hi-Tech District (WND). \textit{(Corresponding authors: Rui Yang, Yutao Yue.)}}
\thanks{Yuanyuan Zhang and Runwei Guan are with the School of Advanced Technology, Xi'an Jiaotong-Liverpool University, Suzhou, 215123, China, the Department of Electrical Engineering and Electronics, University of Liverpool, Liverpool, L69 3GJ, United Kingdom, and also with the Institute of Deep Perception Technology, JITRI, Wuxi, 214000, China (email: Yuanyuan.Zhang16@student.xjtlu.edu.cn; Runwei.Guan21@student.xjtlu.edu.cn).}
\thanks{Lingxiao Li is with the Multimedia Lab (MMLab), Department of Information Engineering, the Chinese University of Hong Kong, Shatin, N.T. 999077, Hong Kong (email: lingxiaoli@cuhk.edu.hk).}
\thanks{Rui Yang and Eng Gee Lim are with the School of Advanced Technology, Xi'an Jiaotong-Liverpool University, Suzhou, 215123, China (email: R.Yang@xjtlu.edu.cn; Enggee.Lim@xjtlu.edu.cn).}
\thanks{Yutao Yue is with the Thrust of Artificial Intelligence and Thrust of Intelligent Transportation, The Hong Kong University of Science and Technology (Guangzhou), Guangzhou 511400, China, and also with the Institute of Deep Perception Technology, JITRI, Wuxi 214000, China. (email: yutaoyue@hkust-gz.edu.cn).}}
% <-this % stops a space

\maketitle

\begin{abstract}
Radar-based cardiac monitoring has become a popular research direction recently, but the fine-grained electrocardiogram (ECG) signal is still hard to reconstruct from millimeter-wave radar signal. The key obstacle is to decouple cardiac activities in the electrical domain (i.e., ECG) from that in the mechanical domain (i.e., heartbeat), and most existing research only uses purely data-driven methods to map such domain transformation as a black box. Therefore, this work first proposes a signal model that considers the fine-grained cardiac feature sensed by radar, and a novel deep learning framework called radarODE is designed to extract both temporal and morphological features for generating ECG. In addition, ordinary differential equations are embedded in radarODE as a decoder to provide morphological prior, helping the convergence of the model training and improving the robustness under body movements. After being validated on the dataset, the proposed radarODE achieves better performance compared with the benchmark in terms of missed detection rate, root mean square error, Pearson correlation coefficient with improvements of $9\%$, $16\%$ and $19\%$, respectively. The validation results imply that radarODE is capable of recovering ECG signals from radar signals with high fidelity and can potentially be implemented in real-life scenarios.
\end{abstract}


\section{Introduction}
Radar-based human sensing is a rapidly evolving field that leverages radio-frequency signals to detect or recognize human activities in many future scenarios (e.g., smart home, in-cabin monitoring and biometric recognition~\cite{hu2023real,islam2022heartprint,wang2022heartprint}). Compared with other contactless sensors such as cameras and acoustic sensors with privacy issues, radar could sense the ambient environment in a non-invasive manner and achieve good robustness under light conditions or temperature variations~\cite{guan2023achelous,luo2023edgeactnet}. In exchange, radar signals cannot be directly interpreted by humans like other mediums with explicit meanings (e.g., images, sound), increasing the complexity of designing efficient frameworks for specific sensing tasks (e.g., pose estimation~\cite{yu2023mobirfpose}, object detection~\cite{mercuri2019vital} and vital sign monitoring~\cite{zhang2023overview}).

Within all scenarios suitable for radar-based human sensing, contactless vital sign monitoring is a crucial task in providing healthcare information (e.g., respiration, heart rate and electrocardiogram (ECG)). The first attempt at radar-based respiration monitoring can be traced back to 1975 by measuring the displacement of the chest wall induced by respiration~\cite{lin1975noninvasive}. The chest wall displacement will modulate the phase component of the emitted radar signal, and the latent respiratory information can be demodulated from the phase variation~\cite{wang2020remote}. Similarly, cardiac activities are small-scale displacements that also cause chest wall displacements, but such small displacements are normally ruined by respiration with orders more amplitude. The follow-up researchers are dedicated to extracting cardiac information in the presence of respiration disturbance and also other common noises, such as random body movement (RBM)~\cite{chen2021movi,zhang2020health}, multi-path or multi-person interference~\cite{mercuri2021enabling,islam2022contactless} and radar self-movement~\cite{da2019theoretical,yang2024isense}. 

In the context of cardiac monitoring, most early studies focused on the recovery of coarse cardiac information, such as heart rate (HR), heart sound and heart rate variability (HRV), from the perspectives of radar front-end design or advanced algorithms design~\cite{zhang2023overview}. For example, some advanced types of radar (e.g., frequency modulated continuous wave (FMCW) radar) are designed to enable high range-resolution or multi-person monitoring~\cite{ha2020contactless}, and some baseband signal-processing algorithms are embedded on the radar platform to realize in-phase/quadrature modulation or accurate phase unwrapping~\cite{obadi2021survey}. In addition, various advanced algorithms are applied by leveraging different intrinsic characteristics of cardiac activities to robustly reconstruct cardiac features. For example, cardiac activities normally reveal strong periodicity in the time domain and have dominant peaks on the spectrum, inspiring periodicity-based methods (e.g., template matching~\cite{lv2018doppler}, hidden Markov model~\cite{xia2021radar}) and spectrum-based methods (e.g., Fourier transform~\cite{nosrati2018accurate}, wavelet transform~\cite{mercuri2019vital}) as two major categories in cardiac feature extraction algorithms.

In recent years, the emergence of commercial radar platforms with high operating frequency (millimeter-wave (mm-wave) radar) encourages researchers to extract fine-grained cardiac features (e.g., ECG and seismocardiography (SCG)) from the radar signal~\cite{zhang2023overview}. SCG signal is measured by the accelerometer mounted on the human chest to measure the mechanical vibrations produced by heartbeats, describing the fine-grained cardiac mechanical activities such as aortic/mitral valve opening/closing and isovolumetric contraction~\cite{cocconcelli2020high}. Although these vibrations are subtle, it is still reasonable to directly map the displacements detected by radar to each fine-grained cardiac mechanical activity using high-resolution radar as proved in~\cite{ha2020contactless}. However, radar-based SCG recovery is not widely investigated compared with ECG, because ECG provides more comprehensive information (e.g., atrial/ventricular depolarization~\cite{khairy2007clinical}) for clinical diagnosis.

To reconstruct ECG from radar signal, the most straightforward approach is to directly sense the variation in the scattered electromagnetic field through frequency shift of mm-wave response, and the ECG signal can then be decoupled from the scattered electromagnetic field based on the dynamic model in the form of partial differential equations deduced from cardiac electrophysiology (i.e., ionic concentration in cardiac cells)~\cite{xu2021cardiacwave}. However, the solutions of the entire model are extremely hard to obtain either numerically or analytically, and the constructed models will be changed with respect to different environments and noises due to the Green's function~\cite{li2019intelligent}, causing difficulty in adapting the model in various real-life scenarios. 

The second approach, which is also the most adopted approach, only uses radar to sense the chest region displacement through the reflected radar signal as in coarse cardiac monitoring, but then the researchers must deal with domain decoupling to transform the measured signal from the mechanical domain to the electrical domain to generate ECG measurement. Intuitively, it is reasonable that mechanical conduction and electrical conduction are highly correlated in describing cardiac activities because the electrical changes in cells trigger heart muscle contraction, whereas such a relationship is called excitation-contraction coupling in electrophysiology and is extremely hard to interpret or model by researchers without biological knowledge~\cite{swift2021stop,orkand1964heart}.

In the literature, the existing studies all leverage deep learning methods to extract latent information from enormous radar/ECG pairs and try to learn domain transformation relying on the extraordinary non-linear mapping ability of the deep neural network~\cite{chen2022contactless,wu2023contactless,li2024radarnet,wang2023ecg,zhao2024airecg}. Although these studies could successfully reconstruct the ECG signal from radar, three drawbacks still need to be improved:
\begin{itemize}
    \item There is no existing signal model with a compact form to describe the domain transformation for radar-based ECG reconstruction.
    \item The purely data-driven method could learn the domain transformation as a black box, but researchers can hardly intervene in the learning process to enhance the characteristic peaks of ECG or explain the intrinsic correspondence in domain transformation.
    \item The well-trained deep learning model is not robust to abrupt noises such as body movement~\cite{chen2022contactless,wang2023vital}, because these noises normally have orders of magnitude higher than cardiac-related vibrations, drowning out subtle features and ruining forward propagation of the deep neural network~\cite{wang2023slprof}.
\end{itemize}

Inspired by the above discussions, this paper aims to design a framework for radar-based ECG reconstruction with ordinary differential equations (ODEs) embedded to provide prior knowledge on domain transformation and resist abrupt noises. The contributions of this research can be concluded as:
\begin{itemize}
    \item This study proposes an ODE-embedded framework called radarODE to produce robust long-term ECG recovery against abrupt noises with the aid of morphological ECG features as the reference.
    \item A signal model is designed to describe the fine-grained cardiac feature sensed by radar, enabling further domain transformation between cardiac mechanical and electrical activities, instead of using purely data-driven approaches without any explanation as in the literature.
    \item Based on the proposed signal model, an ODE-embedded module called single-cycle ECG generator (SCEG) is designed to realize the domain transformation by parameterizing the radar signal into sparse representations and hence generate the morphological ECG features as references to resist noise.
\end{itemize}

The rest of the paper is organized as follows. Section~\ref{sec:bg} introduces the background knowledge required for radar-based ECG reconstruction. Section~\ref{sec:method} explains the proposed model for radar signal and the structure of radarODE framework. Section~\ref{sec:exp} and~\ref{sec:dataset} introduce the public dataset used for validation in this research and then present the results obtained with corresponding comparisons and evaluations. Finally, Section~\ref{sec:conclusions} concludes this paper.

\section{Background}\label{sec:bg}
To understand the model and framework proposed later in this paper, this section will first provide the necessary background about the signal model for radar-based coarse cardiac monitoring and then briefly introduce the relationship between cardiac electrical and mechanical activities.
\subsection{Radar-Based Coarse Cardiac Monitoring}
The vanilla signal model for radar-based cardiac monitoring (e.g., heart rate monitoring) using continuous wave (CW) radar starts from the transmitted signal expressed as
\begin{equation}
s_t(t)=A_t\cdot\cos (2 \pi f t+\theta(t))
\end{equation}
where $A_t$ and $f$ are the amplitude and carrier frequency of the transmitted signal, and $\theta(t)$ is the phase noise from the signal generator with respect to time $t$~\cite{droitcour2004range}. In the ideal case, the radar signal is only reflected by a human at a fixed distance $d_0$ with a varying chest displacement as $x(t)$, and the received signal after propagation time $T_p(t)$ can be derived as
\begin{equation}
s_r(t)=A_r\cdot\cos (2 \pi f (t-T_p(t))+\theta(t-T_p(t)))
\end{equation}
with
\begin{equation}
\begin{aligned}
T_p(t) &= \frac{2d(t)}{c} \\
d(t) &= d_0+x(t)
 \end{aligned}
\end{equation}
where $A_r$ is the amplitude of the received signal, $c$ is the light speed and $2d(t)$ represents the round trip distance of the signal between the transmitter and receiver. Then, the received signal can be expanded as
\begin{equation}
s_r(t)=A_r\cdot\cos (2 \pi ft-\frac{4\pi d_0}{\lambda}-\frac{4\pi x(t)}{\lambda}+\theta(t-\frac{2 d_0}{c}-\frac{2 x(t)}{c}))
\end{equation}
where $\lambda$ is the wavelength that equals to $\frac{c}{f}$. According to~\cite{droitcour2004range,lin2022broadband}, it is safe to eliminate changes in amplitude and phase noise term because the chest displacement is much less than the fixed distance (i.e., $x(t)\ll d_0$). Therefore, the approximate received signal is
\begin{equation}
s_r(t)\approx \cos (2 \pi ft-\frac{4\pi d_0}{\lambda}-\frac{4\pi x(t)}{\lambda}+\theta(t-\frac{2 d_0}{c}))
\end{equation}

The received signal $s_r(t)$ will then pass a local oscillator with a low-pass filter to remove the frequency term, and the resultant baseband signal is
\begin{equation}
s_b(t)=cos(\theta_d+\frac{4\pi x(t)}{\lambda}+\Delta \theta(t))
\end{equation}
with
\begin{equation}
\begin{aligned}
 \theta_d &= \frac{4\pi d_0}{\lambda} + \theta_{0}\\
\Delta \theta (t) &= \theta(t)- \theta(t-\frac{2 d_0}{c})
 \end{aligned}
\end{equation}
where $\theta_d$, $\theta_0$ and $\Delta \theta (t)$ are phase shifts affected by different factors such as $d_0$, signal mixer and antenna, and can be set as constant~\cite{lin2022broadband}. Then, the phase signal unwrapped from the baseband signal is obtained as
\begin{equation}
\phi(t) = \theta_d+\frac{4\pi x(t)}{\lambda}+\Delta \theta(t)
\end{equation}

Finally, the vanilla signal model derived above shows that the chest displacement $x(t)$ is involved in the phase variation of the baseband signal as
\begin{equation}\label{equ:phase}
\Delta\phi(t) = \frac{4\pi x(t)}{\lambda}
\end{equation}

The follow-up researchers have proposed various techniques to improve the accuracy of the unwrapped phase signal variation in (\ref{equ:phase}). For example, the in-phase/quadrature modulation is proposed to solve the null point issue~\cite{droitcour2004range}; the differentiate and cross-multiply algorithm is designed to avoid discontinuity in the unwrapped phase signal~\cite{zhang2023overview}. In addition, chest displacement $x(t)$ is a mixture of cardiac activities, respiration and noises (e.g., RBM~\cite{chen2021movi,zhang2020health}, multi-path or multi-person interference~\cite{mercuri2021enabling,islam2022contactless}). Therefore, enormous advanced algorithms are proposed to decompose cardiac information from $x(t)$, as have been reviewed in~\cite{zhang2023overview}.

\subsection{Radar-Based ECG Monitoring as a Domain Transformation Problem}
\begin{figure}[tb] 
    \centering 
    \includegraphics[width=1\columnwidth]{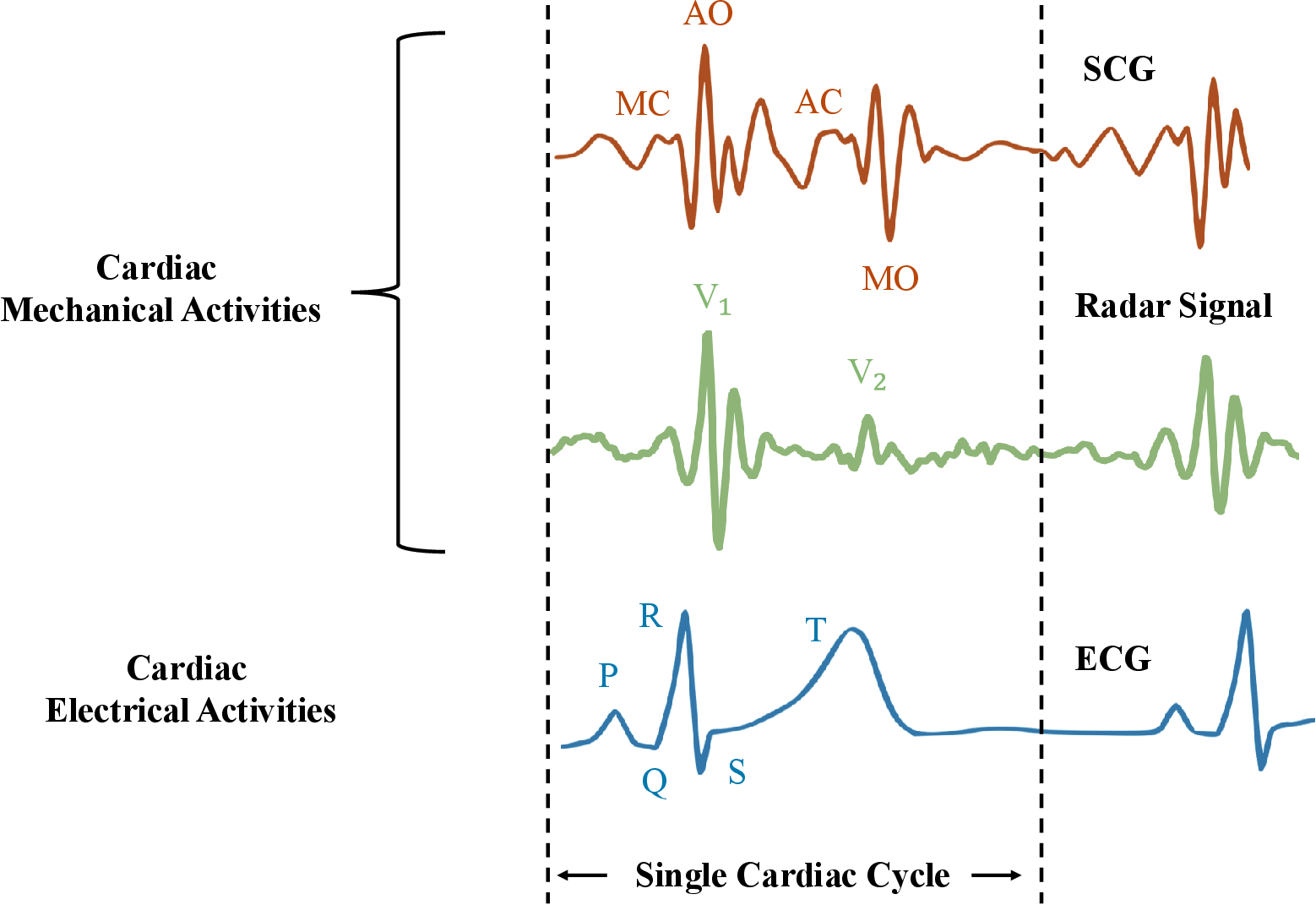}
    \caption{Relationships between cardiac mechanical and electrical activities within the same cardiac cycle.} 
    \label{fig:scg_ecg} 
\end{figure}

Coarse cardiac monitoring only aims to detect a single heartbeat within one cardiac cycle, while fine-grained cardiac monitoring requires recovering subtle cardiac activities within one cardiac cycle. For example, Figure~\ref{fig:scg_ecg} shows the typical radar and SCG signal waveform within a single cardiac cycle that describes the cardiac mechanical activities, such as aortic valve opening/closure (AO/AC) and mitral valve opening/closure (MO/MC)~\cite{swift2021stop,orkand1964heart}. These mechanical activities are muscle contractions stimulated by cardiac electrical events, such as P-wave, QRS-complex and P-wave in the ECG signal as shown in Figure~\ref{fig:scg_ecg}. Therefore, radar-based ECG recovery is actually a domain transformation problem that translates cardiac mechanical activities into electrical activities and realizes fine-grained vital sign monitoring in a contactless manner.
 
In the literature, the domain transformation is only realized by deep-learning-based methods, while a common issue of the deep learning model is not robust against large-scale noise (e.g., RBM) as reported by many previous studies~\cite{wu2023contactless,li2024radarnet,wang2023ecg,zhao2024airecg}. However, there is no existing work that investigates the noise-robustness of the deep learning model itself, and this study is motivated to develop a noise-robust deep learning model to realize the domain transformation in the presence of body movement noise.

\section{Methodology}\label{sec:method}
\subsection{Overview}
\begin{figure*}[tb] 
    \centering 
    \includegraphics[width=2\columnwidth]{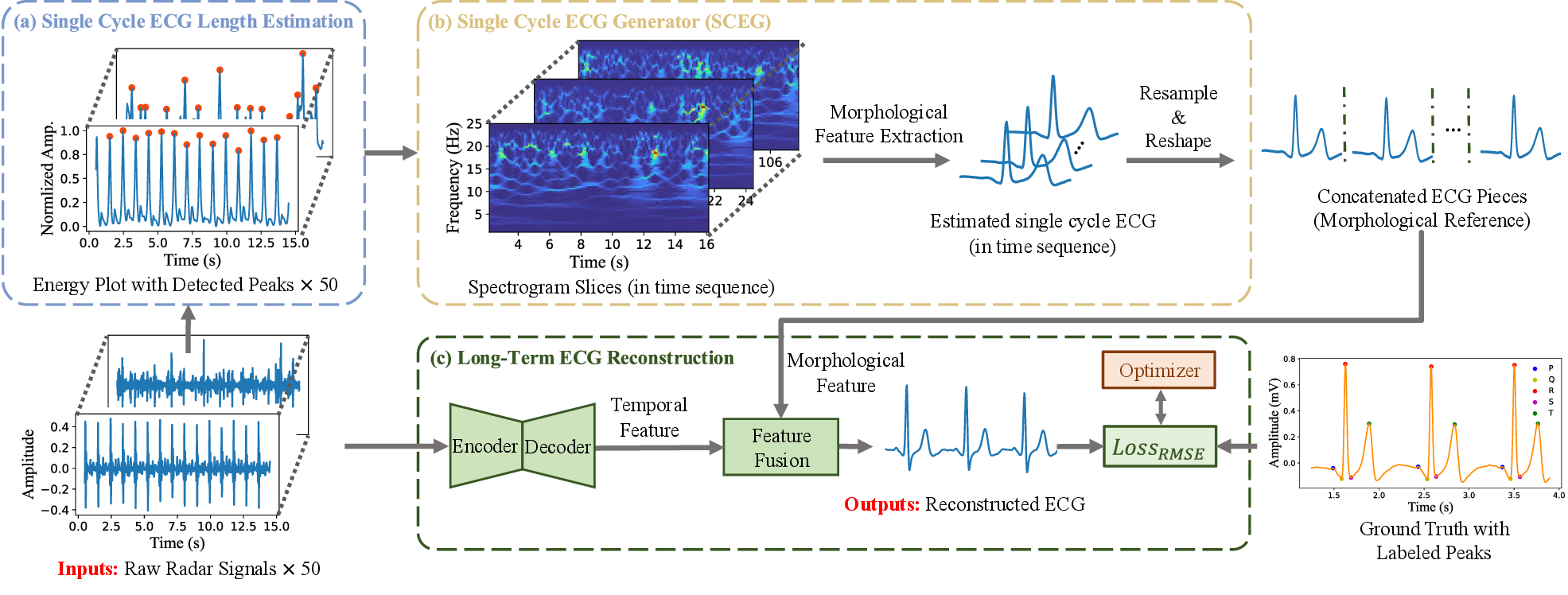}
    \caption{Overview of the radarODE framework for domain transformation: (a) Single-cycle ECG length (PPI) estimation; (b) Single-cycle ECG generator; (c) Long-term ECG reconstruction.}
    \label{fig:radarODE} 
\end{figure*}

In order to realize a robust radar-to-ECG reconstruction, the proposed radarODE framework is designed based on the domain transformation within a single cardiac cycle to generate faithful ECG reference that aids the domain transformation in long-term ECG recovery. The inputs of radarODE are $50$ synchronous radar signals that represent the measurements from $50$ spatial points within the chest region~\cite{chen2022contactless}, as shown in Figure~\ref{fig:radarODE}, and the entire domain transformation is realized by three modules:

\begin{itemize}
    \item The first module estimates the peak-to-peak interval (PPI) for consecutive heartbeats based on the proposed peak detection algorithm and provides information for single cardiac cycle segmentation, as shown in Figure~\ref{fig:radarODE}(a).
    \item The second module first transforms the time-domain radar signal within a single cardiac cycle into a spectrogram using synchrosqueezed wavelet transform (SST). Then, the temporal and ODE decoder within SCEG will generate the faithful ECG pieces as morphological references while resisting noise disruption, as shown in Figure~\ref{fig:radarODE}(b).
    \item The third module is used to fuse the extracted morphological and temporal features hidden in the original radar signal to generate the final ECG recovery, as shown in Figure~\ref{fig:radarODE}(c).
\end{itemize}
  
\subsection{Model for Radar Signal and Pre-Processing}
\subsubsection{Fine-Grained Model for Radar Signal}
According to the discussion in the last section, the chest displacement $x(t)$ can be further decomposed as
\begin{equation}
x(t) = x_c(t) + x_r(t) + x_n(t)
\end{equation}
where $x_c(t)$ means cardiac mechanical activities, $x_r(t)$ is respiration induced displacement and $x_n(t)$ is noise term. After the pre-processing, the respiration term has been filtered out, and the actual radar signal $\tilde{x}(t)$ provided in the dataset~\cite{chen2022contactless} can be expressed as 
\begin{equation}
\tilde{x}(t) = x_c(t) + x_n(t)
\end{equation}

Furthermore, the pre-processed radar signal $\tilde{x}(t)$ has two prominent vibrations $v_1$ and $v_2$ as shown in Figure~\ref{fig:scg_ecg}, corresponding to the fine-grained cardiac mechanical activities shown in SCG. According to the previous work on SCG modeling, the heart muscle contraction has a pulsatile nature, and the bones/tissues in chest area introduce the extra damping into the pulse~\cite{nosrati2018accurate}. Inspired by the natural characteristics, the radar signal with two prominent vibrations measured in a single cardiac cycle is innovatively modeled as the Gaussian pulses with certain central frequencies as
\begin{equation}\label{equ:vib}
\tilde{x}(t) = v_1(t) + v_2(t) + x_n(t)
\end{equation}
with
\begin{equation}
\begin{aligned}
 v_1 &= \mathrm{a}_1 \mathrm{cos}(2\pi f_1 t)\exp \left(-\frac{(t-T_1)^2}{{b_1}^2}\right)\\
v_2 &= \mathrm{a}_2 \mathrm{cos}(2\pi f_2 t) \exp \left(-\frac{(t-T_2)^2}{{b_1}^2}\right)
 \end{aligned}
\end{equation}
where $a_1$, $b_1$ and $a_2$, $b_2$ jointly contribute to the amplitudes and length of the first and second prominent vibrations, $f_1$, $f_2$ are the corresponding central frequencies, $T_1$, $T_2$ determine when the vibrations happen, and $x_{n}(t)$ represents all the noises. 

The aim of proposing the model in (\ref{equ:vib}) is not to perform the curve fitting but to provide the explanation for the later radarODE design, because the positions of the prominent vibrations ($T_1$, $T_2$) are crucial to the precise reconstruction of ECG peaks using deep neural network.

\subsubsection{Signal Pre-Processing with Synchrosqueezed Wavelet Transform (SST)}
Based on the proposed radar signal model in (\ref{equ:vib}), the next step is to enhance the prominent vibrations (i.e., $T_1$, $T_2$). Figure~\ref{fig:scg_ecg} shows that the high SNR radar signal could reveal prominent peaks of $v_1$ and $v_2$ in the time domain, but in most cases, these two peaks (especially $v_2$) could be ruined by noise. Therefore, this research decides to extract the time-frequency domain information from the spectrogram obtained by synchrosqueezed wavelet transform (SST)~\cite{daubechies2011synchrosqueezed}, and the two vibrations can then be localized by the SCEG module proposed later in Section~\ref{sec:sceg}. 

SST evolves from continuous wavelet transform (CWT) but with concentrated energy distribution along the frequency axis, providing a sparser time-frequency representation with enhanced prominent vibrations compared with other tools such as short-time Fourier transform (STFT) and CWT.

The first step of SST is to calculate the CWT of radar signal $\tilde{x}(t)$ as
\begin{equation}
W_{\tilde{x}}(a, b)=\int \tilde{x}(t) a^{-1 / 2} \psi^*\left(\frac{t-b}{a}\right) d t
\end{equation}
where $\psi^*$ is the complex conjugate of the chosen mother wavelet, and $a$, $b$ are the adjustable scaling and translation factors for the wavelet $\psi$ to extract frequency- and time-domain information, respectively. In this research, the Morlet wavelet is selected as mother wavelet because it is widely used for vibration signal processing, especially for time-frequency localization~\cite{xue2023novel}.

The second step is to calculate the candidate instantaneous frequency for $W_{\tilde{x}}(a, b)\neq 0$ as
\begin{equation}
f_{\tilde{x}}(a, b)=-2\pi i\left(W_{\tilde{x}}(a, b)\right)^{-1} \frac{\partial W_{\tilde{x}}(a, b)}{\partial b}
\end{equation}

The final step is to concentrate the energy along the candidate instantaneous frequency as
\begin{equation}
T_{\tilde{x}}(2\pi f, b)=\int_{A(b)} W_{\tilde{x}}(a, b) a^{-3 / 2} \delta\left(2\pi f_{\tilde{x}}(a, b)-2\pi f\right) d f
\end{equation}
where $A(b)=\left\{a ; W_{\tilde{x}}(a, b) \neq 0\right\}$, and $\delta$ represents the Dirac-delta function in a distribution version to smoothly squeeze the spread-out energy into a narrow band around the instantaneous frequency~\cite{daubechies2011synchrosqueezed}.

\begin{figure}[tb]
        \centering
        \subfloat[]{\label{fig:compare_radar}\includegraphics[width=0.5\columnwidth]{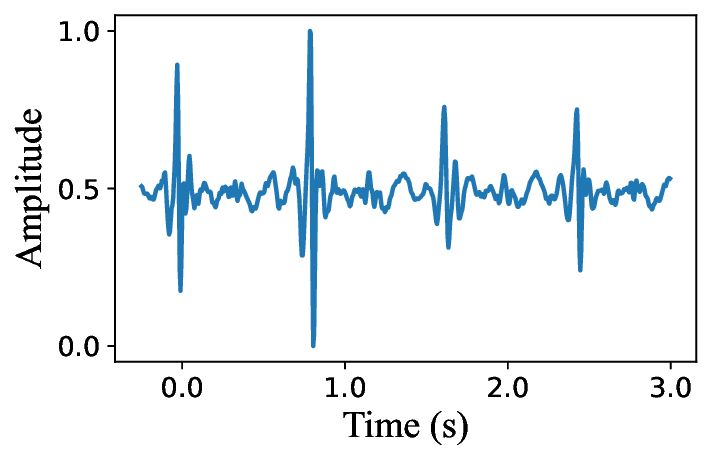}}
        \subfloat[]{\label{fig:compare_stft}\includegraphics[width=0.5\columnwidth]{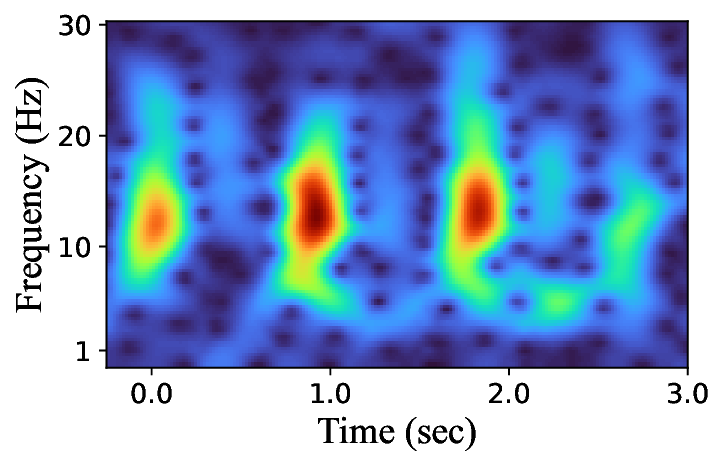}} \\ \vspace{-0.04\columnwidth}
        \subfloat[]{\label{fig:compare_cwt}\includegraphics[width=0.5\columnwidth]{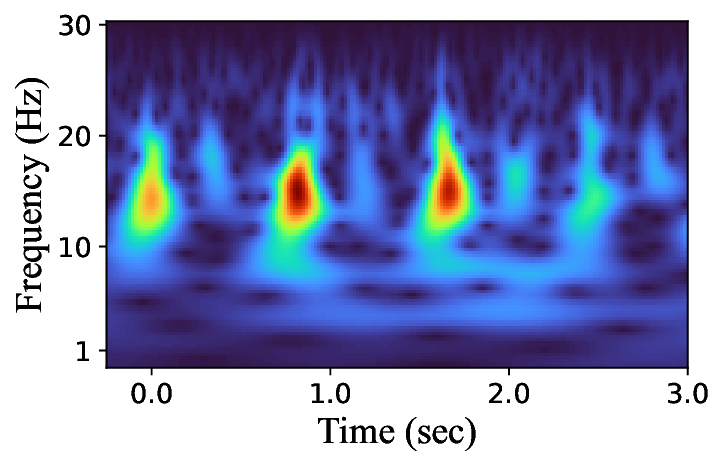}}
        \subfloat[]{\label{fig:compare_sst}\includegraphics[width=0.5\columnwidth]{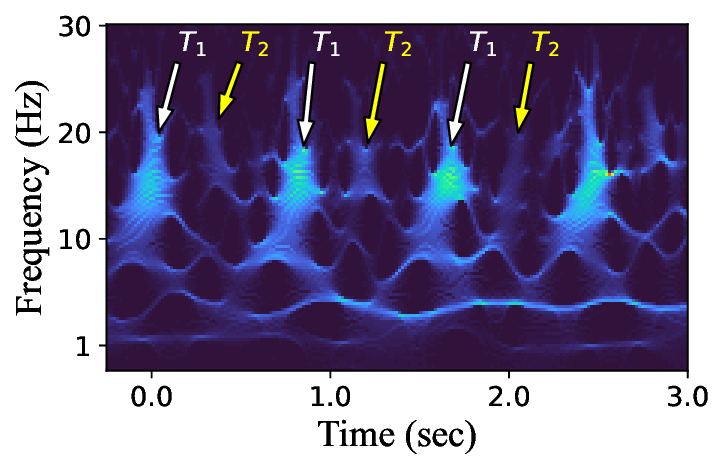}} 
        \caption{Spectrograms obtained from the same radar signal: (a) Radar signal $\tilde{x}(t)$; (b) STFT result; (c) CWT result; (d) SST result with $T_1$, $T_2$ labeled, revealing concentrated energy around vibrations and clean background.}
        \label{fig:spec_compare}
\end{figure}

The quality of the resultant spectrograms can be evaluated using power spectrogram entropy (PSE)~\cite{wang2023vital}, with a small value indicating that the energy is concentrated around a certain frequency. The calculated PSE for the spectrogram produced by STFT, CWT and SST is $0.94$, $0.90$ and $0.76$ respectively, with the visualized results shown in Figure~\subref*{fig:compare_stft},~\subref*{fig:compare_cwt} and~\subref*{fig:compare_sst}. It is clear that the spectrogram obtained by STFT only shows the rough positions of each $v_1$, while CWT gives a sharp position for both vibrations, but the energy is still spread out. By further concentrating the energy distribution, SST produces the spectrogram with a relatively clean background and sharp peaks for the vibrations, reducing the burden of the deep-learning-based SCEG module in extracting latent features (i.e., $T_1$, $T_2$).

\subsection{radarODE Framework Design}
\subsubsection{Single-cycle ECG Length Estimation}
\begin{figure}[tb]
        \centering
        \subfloat[]{\label{fig:correct_detection}\includegraphics[width=0.5\columnwidth]{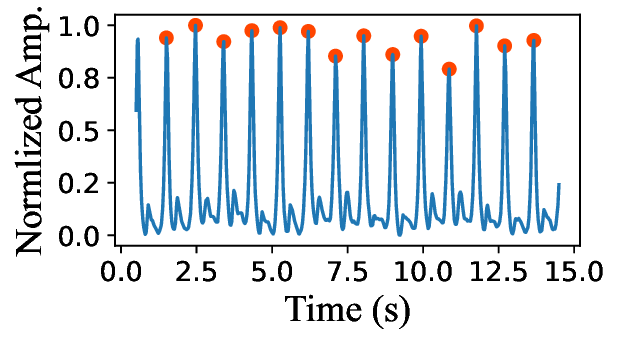}}
        \subfloat[]{\label{fig:wrong_detection}\includegraphics[width=0.5\columnwidth]{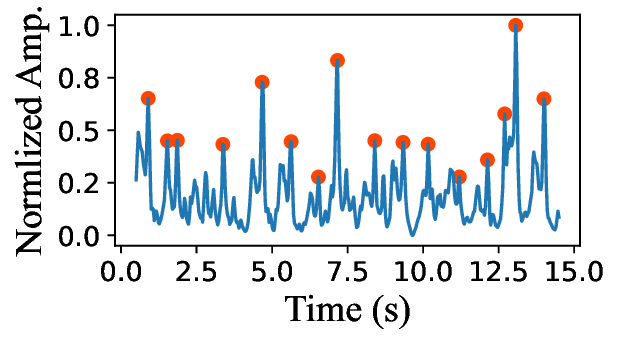}}
        \caption{Energy plot of the synchronous radar signals with different detected peaks: (a) Energy plot with high SNR and correct detection; (b) Energy plot with low SNR and wrong detection.}
        \label{fig:detection_compare}
\end{figure}

The first module of radarODE aims to estimate the length of each single-cycle ECG piece by calculating the interval between two consecutive heartbeats (i.e., PPI) from the energy plots $\hat{x}$ of the radar signal $\tilde{x}$ (omit $(t)$ for simplicity) as shown in Figure~\subref*{fig:correct_detection}. The energy plot is obtained by simply adding the spectrogram along frequency axis, but the peak detection results may not be promising due to low SNR signals as shown in Figure~\subref*{fig:wrong_detection}. Therefore, a new algorithm is proposed as in Algorithm~\ref{alg:ppi} to eliminate the wrong detection obtained from $50$ radar energy plots ${X}_{\mathcal{L}}$, with the length of each $\hat{x}_{i}\in {X}_{\mathcal{L}}$ equal to $l$.

The design of Algorithm~\ref{alg:ppi} is based on the fact that the PPI for healthy people tends to be unchanged in adjacent cardiac cycles. In this case, the long-term radar energy plots are firstly sliced into short segments as shown in the \textsc{Initialization} stage in Algorithm~\ref{alg:ppi}, and then the biopeaks algorithm implemented in NeuroKit2~\cite{makowski2021neurokit2} is used for detecting all the potential peaks $P$ from each energy plot segment $\hat{x}^j_i$ as:
\begin{equation}\label{equ:biopeak}
P = biopeaks(\hat{x}^j_i)
\end{equation}
Secondly, the resultant $PPI_\mathcal{C}$ obtained from Line~\ref{line:biopeaks}-\ref{line:update} in Algorithm~\ref{alg:ppi} contains potential estimated PPI from $50$ radar energy plots, with the correct estimations as the majority. Therefore, the kernel density estimation (KDE)~\cite{terrell1992variable} is applied on the candidate set $PPI_\mathcal{C}$ to calculate the probability density of different PPI values as:
\begin{equation}\label{equ:kde}
KDE:\ \hat{f}(p)=\frac{1}{n h} \sum_{c=1}^n K\left(\frac{p-PPI_{c\in \mathcal{C}}}{h}\right)
\end{equation}
where $\hat{f}(p)$ means the estimated probability density function at point $p$, $n$ is the number of all the estimated PPI in $PPI_\mathcal{C}$, $K$ is the Gaussian kernel function and $h=n^{-1 /5}$ is the bandwidth of the kernel. Lastly, the final PPI for the current segment is selected as the argument $p$ when $\hat{f}(p)$ achieves the maximum as in Line~\ref{line:final} in Algorithm~\ref{alg:ppi}, and the long-term PPI estimation can be obtained step by step as Algorithm~\ref{alg:ppi} terminated.

\begin{algorithm}[tb]
\caption{PPI Estimation}\label{alg:ppi}
\begin{algorithmic}[1]
    \State \textbf{Input:} \multiline{0.4}{Radar Energy Plots ${X}_{\mathcal{L}} = \{\hat{x}_{1},\hat{x}_{2},\ldots,\hat{x}_{50}\}$, \\ Segments Length $l_{seg}$ and Step Length $l_{step}$}
    \State \textbf{Output:} Estimated $PPI$
    \Statex \textsc{Initialization}:
    \State\quad - \multiline{0.4}{Let ${X}_{\mathcal{S}}=\{{X}^1_{\mathcal{L}}, {X}^2_{\mathcal{L}},\ldots, {X}^{J}_{\mathcal{L}}\}$ be an ordered list of the segment lists sliced from ${X}_{\mathcal{L}}$ with length $l_{seg}$ and step $l_{step}$, where $J = \frac{l-l_{seg}}{l_{step}}$.}
    \State\quad - Let $PPI \leftarrow \emptyset$.
    \Statex \textsc{Main iteration}:
    \For{each segment list ${X}^j_{\mathcal{L}}\in {X}_{\mathcal{S}}$}
        \State - \multiline{0.4}{Let $PPI_{\mathcal{C}} \leftarrow \emptyset$ to save the candidate PPI obtained from each segment.}
        \For{each segment $\hat{x}^{j}_{i} \in {X}^j_{\mathcal{L}}$}
            \State 1) \multiline{0.4}{ Apply biopeaks~\cite{makowski2021neurokit2} on $\hat{x}^{j}_{i}$ to get all the detected peaks $P$ as in (\ref{equ:biopeak}).} \label{line:biopeaks}
            \State 2) \multiline{0.4}{ Get PPI for the current radar signal segment using differentiation as $PPI_c \leftarrow diff(P$).}
            \State 3)  Update $PPI_{\mathcal{C}} \leftarrow PPI_{\mathcal{C}} \cup PPI_c$. \label{line:update}
        \EndFor
    \State - \multiline{0.4}{Calculate the probability density function $\hat{f}(p)$ for $PPI_{\mathcal{C}}$ using KDE as in (\ref{equ:kde}).}
    \State - \multiline{0.4}{Determine the final PPI for the current segment list and update the set as $PPI \leftarrow PPI \cup\arg \max_{p} \hat{f}(p)$.} \label{line:final}
    \EndFor
\end{algorithmic} 
\end{algorithm} 

\subsubsection{Single-cycle ECG Generator (SCEG)}\label{sec:sceg}
Based on the yielded $PPI$, the SST spectrogram can be sliced into segments corresponding to a single cardiac cycle, and the aim of the SCEG module is to reconstruct the ECG for each single cardiac cycle, hence realizing the transformation from mechanical to electrical domain. In general, the input of the SCEG is $N$ segments of the SST plot within $[1,25]$~Hz with the size of $F\times T$ on frequency and time axis, and the output is the corresponding $N$ ECG pieces with the same length $T$, as shown in Figure~\ref{fig:sceg}. In practice, the deep neural network only accepts the inputs/outputs of the same size. Therefore, the actual SST segment is centered at the current cardiac cycle and expands to $4$ seconds, and the corresponding ECG ground truth is resampled to a fixed length of $200$ for loss calculation.

For architecture design, the SCEG module adopts the popular backbone-encoder-decoder structure as verified by enormous image-related tasks~\cite{li2023euclidean,chu2024vessel}, with detailed parameters shown in Table~\ref{tab:param}. In addition, a feature fusion block is added after the decoder to fuse the temporal and morphological features and generate the final ECG reconstructions. The detailed implementation of each part in  Table~\ref{tab:param} with explanations can be elaborated as:

\begin{figure*}[tb] 
    \centering 
    \includegraphics[width=2\columnwidth]{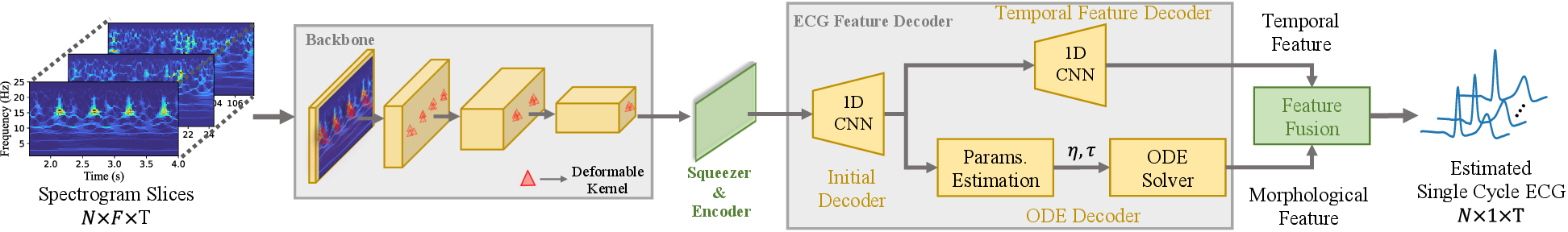}
    \caption{Architecture of SCEG with SST segments as input and single-cycle ECG pieces as output.} 
    \label{fig:sceg} 
\end{figure*}

\begin{table}[tb]
\centering
\caption{Structure and Parameters for SCEG}
    \begin{tabular}{l l l}
    \toprule
    \textbf{Layers} & \textbf{Parameters}  & \textbf{Output Shape} \\
    & ($C_{in}$, $C_{out}$, $K$, $S$)$^1$ & $N$: Batch Size \\
    \toprule
    Input SST &  & $(N, 50, 71, 118)$ \\
    \toprule
    \textbf{a. Backbone} & & \\
    \ \ Residual Block   & $(50,128,(2,1),(1,1))$  & $(N, 128, 72, 118)$\\
    \ \ Downsample Block & $(128,128,(3,2),(2,2))$ &  $(N, 128, 36, 60)$\\
    \ \ Residual Block   & $(128,256,(2,1),(1,1))$ & $(N, 256, 37, 60)$\\
    \ \ Downsample Block & $(256,256,(3,2),(2,2))$ &  $(N, 256, 19, 31)$\\
    \ \ Residual Block   & $(256,512,(2,1),(1,1))$ & $(N, 512, 20, 31)$\\
    \ \ Downsample Block & $(512,512,(3,3),(2,1))$ &  $(N, 512, 10, 31)$\\
    \ \ Residual Block   & $(512,1024,(2,1),(1,1))$ & $(N, 1024, 11, 31)$\\
    \ \ Downsample Block & $(1024,1024,(3,3),(2,1))$ &  $(N, 1024, 6, 31)$\\
    \toprule
    \textbf{b. Squeezer\&Encoder} & & \\
    \ \ Conv2d & $(1024,1024,(6,1),(1,1))$  & $(N, 1024, 31)$\\
    \ \ Transconv1d Block & $(1024,512,5,3)$ & $(N, 512,95)$ \\
    \ \ Transconv1d Block & $(512,256,5,3)$ & $(N, 256,287)$ \\
    \ \ Transconv1d Block & $(512,128,5,3)$ & $(N, 128, 863)$ \\
    \toprule
    \multicolumn{3}{l}{\textbf{c. ECG Feature Decoder}} \\
    \multicolumn{3}{l}{\ \ \textbf{Initial Decoder}} \\
    \ \ Conv1d Block  & $(128,64,7,2)$ & $(N,64,430)$ \\
    \ \ Conv1d Block  & $(64,32,7,2)$ & $(N,32,213)$ \\
    \ \ Conv1d Block  & $(32,16,7,1)$ & $(N,16,209)$ \\
    \ \ Conv1d Block  & $(16,8,5,1)$ & $(N,8,207)$ \\
    \multicolumn{3}{l}{\ \ \textbf{Temporal Feature Decoder}} \\
    \ \ Conv1d  & $(8,4,7,1)$ & $(N,4,203)$ \\
    \ \ Conv1d  & $(4,2,5,1)$ & $(N,2,201)$ \\
    \ \ Conv1d  & $(2,1,2,1)$ & $(N,1,200)$ \\
    \ \ \textbf{ODE Decoder} & \\
    \ \ Linear Block & $(8*207, 512,-,-)$ & $(N, 512)$  \\
    \ \ Linear Block & $(512, 128,-,-)$ & $(N, 128)$  \\
    \ \ Linear Block & $(128, 32,-,-)$ & $(N, 32)$ \\
    \ \ Linear Block & $(32, 16,-,-)$ & $(N, 16)$  \\
    \ \ ODE Solver   & $-$ & $(N,1,200)$ \\
    \toprule
    \textbf{d. Feature Fusion} & &  \\
    \ \ Feature Multiply  &  $-$ & $(N,1,200)$ \\
    \ \ Stack  &  $-$ & $(N,1,4,200)$ \\
    \ \ Conv2d Block  & $(1,16,(5,5),(1,2))$ & $(N,16,2,98)$ \\
    \ \ Conv2d Block  & $(16,32,(3,3),(1,2))$ & $(N,32,2,48)$ \\
    \ \ Conv2d Block  & $(32,64,(3,3),(2,2))$ & $(N,64,1,23)$ \\
    \ \ Transconv1d Block  & $(64,32,5,2)$ & $(N,32,52)$ \\
    \ \ Transconv1d Block  & $(32,16,5,2)$ & $(N,16,106)$ \\
    \ \ Transconv1d Block  & $(16,8,3,2)$ & $(N,8,10,211)$ \\
    \ \ Transconv1d  & $(8,4,6,1)$ & $(N,4,206)$ \\
    \ \ Transconv1d  & $(4,2,5,1)$ & $(N,2,202)$ \\
    \ \ Transconv1d  & $(2,1,3,1)$ & $(N,1,200)$ \\
    \bottomrule
    \multicolumn{2}{l}{Output single-cycle ECG piece} &  $(N, 1, 200)$ \\
    \bottomrule
    \multicolumn{3}{l}{1. $C_{in}$: Input channel,  $C_{out}$: Output channel, $K$: Kernel size, $S$: Stride}\\
    \end{tabular}%
\label{tab:param}%
\end{table}%
\begin{enumerate}[a.]
\item \textbf{Backbone: } Backbone is typically used as the first block to extract both low-level (e.g., color, edge) and high-level features (e.g., presence of specific pattern) from the input images. In the context of this research, the backbone is expected to localize the vibrations $v_1$, $v_2$ revealed as periodically appeared bright triangles within the range of $[1,25]~Hz$ on SST plots, providing latent information of $T_1$, $T_2$ for the further ECG reconstruction. 

\quad According to the literature, ResNet with residual blocks is widely used as the backbone for feature extraction in many fields~\cite{chen2024tfpred}, and this research will use a similar structure with $4$ layers of residual blocks and downsample blocks as shown in Figure~\ref{fig:backbone}, with the key parameters listed in Table~\ref{tab:param}. In addition, the traditional 2D convolution is all replaced by deformable 2D convolution~\cite{dai2017deformable} with deformable kernels (instead of square kernels) to fit the irregular shape of the target patterns, as shown in red triangles in Figure~\ref{fig:sceg}.
\begin{figure}[tb] 
    \centering 
    \includegraphics[width=0.8\columnwidth]{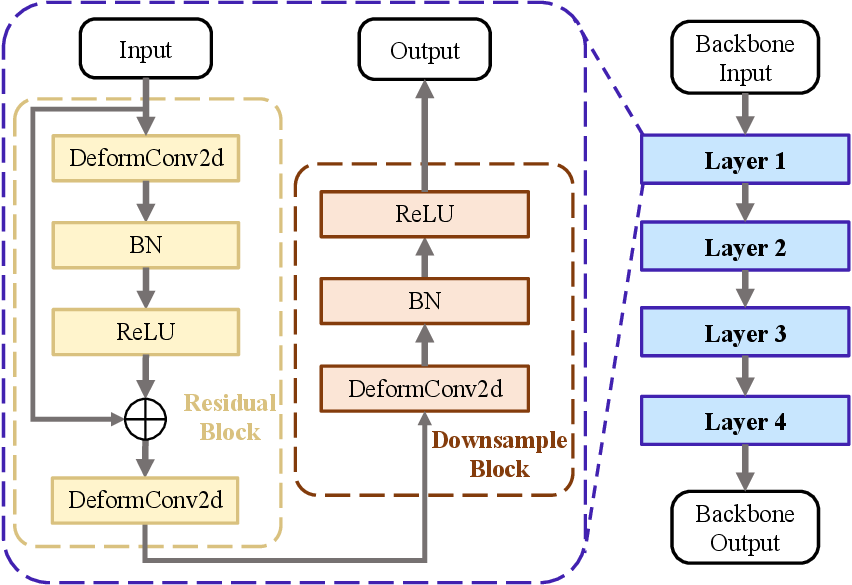}
    \caption{Structure of backbone composed of residual block and downsample block: DeformConv2d means deformable 2D convolution, BN is batch normalization, and ReLU is the rectified linear unit activation function.} 
    \label{fig:backbone} 
\end{figure}

\item \textbf{Squeezer and Encoder: } The output feature map from the backbone should involve both frequency- and time-domain features, and the squeezer is simply used to squeeze the frequency-domain feature using 2D convolution (Conv2d) and output the feature map with the temporal feature only. Then, the encoder assembled by three 1D transposed convolution (Transconv1d) blocks is used to further extract the temporal feature, with each block comprising Transconv1d, BN and ReLU. Finally, the output feature map from the encoder should contain latent information about $v_1$, $v_2$ (i.e., ($T_1$, $T_2$)).

\item \textbf{ECG Feature Decoder: }The decoder is an essential part of SCEG to extract the temporal and morphological features separately from the latent information in the SST feature map, as shown in two branches in Figure~\ref{fig:sceg}. At first, an initial decoder is shared by the latter two decoders and comprises four 1D convolution (Conv1d) blocks with Conv1d, BN, and ReLU inside. Similarly, the temporal feature decoder is assembled by three Conv1d, and the output feature should contain prominent peaks at the position of $T_1$, $T_2$. However, these two peaks may still have deviations from the peaks in ECG ground truth, because obviously the mechanical vibrations $v_1$, $v_2$ lag behind QRS-complex and T-peaks, as shown in Figure~\ref{fig:scg_ecg}. Another problem with the temporal feature decoder is that the ECG pieces have an entirely different shape with radar measurements, and the decoder needs to `memorize' the unique pattern of ECG. Although the previous work shows the deep neural network could learn the patterns after training, the whole process of ECG reconstruction lacks supervision and is vulnerable to noise in radar signals~\cite{chen2022contactless,yang2022pipeline}.

\begin{table}[tb]
\centering
\caption{Default values for $\eta$}
    \begin{tabular}{cccccc}
    \toprule
    \diagbox{$\mathbf{\eta}$}{$\mathbf{\mathcal{F}}$} & $\mathbf{e_P}$ &  $\mathbf{e_Q}$& $\mathbf{e_R}$ & $\mathbf{e_S}$ & $\mathbf{e_T}$\\
    \toprule
    $a_{e_f}$ & $5$ & $-100$ & $480$ & $-120$ & $8$ \\
    $b_{e_f}$ & $0.25$ & $0.1$ & $0.1$ & $0.1$ & $0.4$ \\
    $\theta_{e_f}$ & $\frac{-15\pi}{180}$ & $\frac{25\pi}{180}$ & $\frac{40\pi}{180}$ & $\frac{60\pi}{180}$ & $\frac{135\pi}{180}$ \\
    \bottomrule
    \end{tabular}%
\label{tab:ode_value}%
\end{table}%

In this case, the ODE decoder is designed as a branch to assist the transformation between cardiac mechanical and electrical activities. The main obstacles in modeling such domain transformation are the lack of (a) A compact model for ECG signal and (b) A corresponding explanation between parameters in describing radar signal and ECG signal, e.g., what is the relationship between R peak and $v_1$ in Figure~\ref{fig:scg_ecg}. In this work, the aforementioned two obstacles can be solved from the following perspectives:
\begin{itemize}
    \item The shape of the ECG piece can be modeled morphologically using ODEs in a compact form without any biological/chemical knowledge~\cite{mcsharry2003dynamical}.
    \item The measurements of mechanical activities generally lag behind those of electrical activities with a short time delay $\tau$~\cite{swift2021stop,orkand1964heart}.
\end{itemize}

\quad Inspired by the above facts, the ODE decoder is designed as a parameter estimation part and an ODE solver as shown in Figure~\ref{fig:sceg}, and the solution of the ODEs will be shifted to the left with time $\tau$. In this manner, the latent information in describing radar signals is first transformed to the parameters for the ECG signal, and then the ODE solver will generate morphological-prior to accelerate the convergence of the model training process and provide extra robustness against noises.

\quad To be specific, the parameters estimation part contains four linear blocks (Linear Layer, BN, Tanh) to project the latent space yielded by the initial decoder into parameters $\eta$, $\tau$, and $\eta$ will be sent to an ODE solver to solve a 3D trajectory denoted by $(x,\ y,\ z)$ as
\begin{equation}\label{equ:ode}
    \left\{
    \begin{aligned}
    & \frac{d x}{d t}=\alpha(x, y) x-\omega y \\
    & \frac{d y}{d t}=\alpha(x, y) y+\omega x \\
    & \frac{d z}{d t}=-\sum_{e_f\in \mathcal{F}} a_{e_f} \Delta \theta_{e_f}(x, y) e^{-\Delta \theta_{e_f}(x, y)^2 / 2 b_{e_f}^2}-z
    \end{aligned}
    \right. 
\end{equation}
    with
    \begin{equation}
    \begin{aligned}
    \alpha(x, y) & =1-\sqrt{x^2+y^2} \\
    \Delta \theta_{e_f}(x, y) & =\left(\theta(x, y)-\theta_{e_f}\right) \quad \bmod 2 \pi \\
    \theta(x, y) & =\operatorname{atan} 2(y, x)) \in[-\pi, \pi] \\
    e_f\in\mathcal{F} & =\{e_P, e_Q, e_R, e_S, e_T\}
    \end{aligned}
\end{equation}
where $\mathcal{F}$ represents five characteristic peaks (PQRST) in a single-cycle ECG signal, and the whole ODEs can be interpreted as manipulating each peak along a unit circle by varying the value of $\eta=\{a_{e_f},b_{e_f},\theta_{e_f}\}$ to adjust corresponding amplitude, width and position of each peak. After specifying all $15$ parameters $\eta$ ($3$ for each peak) and the initial conditions of $(x,y,z)$, the value of $z$ can be solved by the ODE solver using the Euler method to get the final single-cycle ECG signal as the morphological feature. In practice, the default values for $\eta$ are provided in advance as in Table~\ref{tab:ode_value}, and the estimated parameters within the range of $[-1,1]$ are used to scale the default values.

\item \textbf{Feature Fusion: } The feature fusion module could leverage respective advantages of the morphological and temporal features and generate the final ECG signal for loss calculation, because the morphological feature only focuses on five peaks to provide a rough shape of ECG with calibrated peaks, and the temporal feature reserves all the other feature neglected in (\ref{equ:ode}) to help the final reconstruction~\cite{guan2023achelous}. Therefore, two features are first fused together by multiplication (Mul.) into one and then stacked four times by itself. Then, the stacked feature is encoded and decoded as in Table~\ref{tab:param} to produce the final single-cycle ECG piece.
\end{enumerate}

The last step of SCEG is to resample all the ECG pieces generated after the feature fusion part with respect to the previous PPI estimation, and the resampled ECG pieces are concatenated in a time sequence to form the final morphological reference for long-term ECG reconstruction.

\begin{table}[tb]
\centering
\caption{Structure and Parameters for Long-term ECG Reconstruction}
    \begin{tabular}{l l l}
    \toprule
    \textbf{Layers} & \textbf{Parameters}  & \textbf{Output Shape} \\
    & ($C_{in}$, $C_{out}$, $K$, $S$)$^1$ & $N$: Batch Size \\
    \toprule
    Input raw radar Signal &  & $(N, 50, 800)$ \\
    \toprule
    \textbf{a. Encoder} & & \\
    \ \ Residual Block   & $(50,128,5,1)$  & $(N, 128, 800)$\\
    \ \ Downsample Block & $(128,128,5,2)$ &  $(N, 128, 400)$\\
    \ \ Residual Block   & $(128,256,5,1)$ & $(N, 256, 400)$\\
    \ \ Downsample Block & $(256,256,5,2)$ &  $(N, 256, 200)$\\
    \ \ Residual Block   & $(256,512,5,1)$ & $(N, 512, 200)$\\
    \ \ Downsample Block & $(512,512,5,2)$ &  $(N, 512, 100)$\\
    \toprule
    \textbf{b. Decoder} & & \\
    \ \ Transconv1d Block & $(512,128,5,2)$ & $(N, 128,200)$ \\
    \ \ Transconv1d Block & $(128,16,5,2)$ & $(N, 16,400)$ \\
    \ \ Transconv1d Block & $(16,1,5,2)$ & $(N, 1, 800)$ \\
    \toprule
    \multicolumn{2}{l}{\textbf{c. Feature Fusion (TCN)}}  \\
    \ \ Feature Stack$^2$  &  $-$ & $(N,2,800)$ \\
    \ \ Dilated Conv1d $\times 9$  & $K=3$, $D = 2$ & $(N, 1, 800)$ \\
    \bottomrule
    \multicolumn{2}{l}{Output long-term ECG} &  $(N, 1, 800)$ \\
    \bottomrule
    \multicolumn{3}{l}{1. $C_{in}$: Input channel,  $C_{out}$: Output channel, $K$: Kernel size, $S$: Stride}\\
    \multicolumn{3}{l}{2. Stack with the morphological feature as in Figure~\ref{fig:radarODE}(c).}\\
    \end{tabular}%
\label{tab:long_structure}
\end{table}%

\begin{figure}[tb] 
    \centering 
    \includegraphics[width=1\columnwidth]{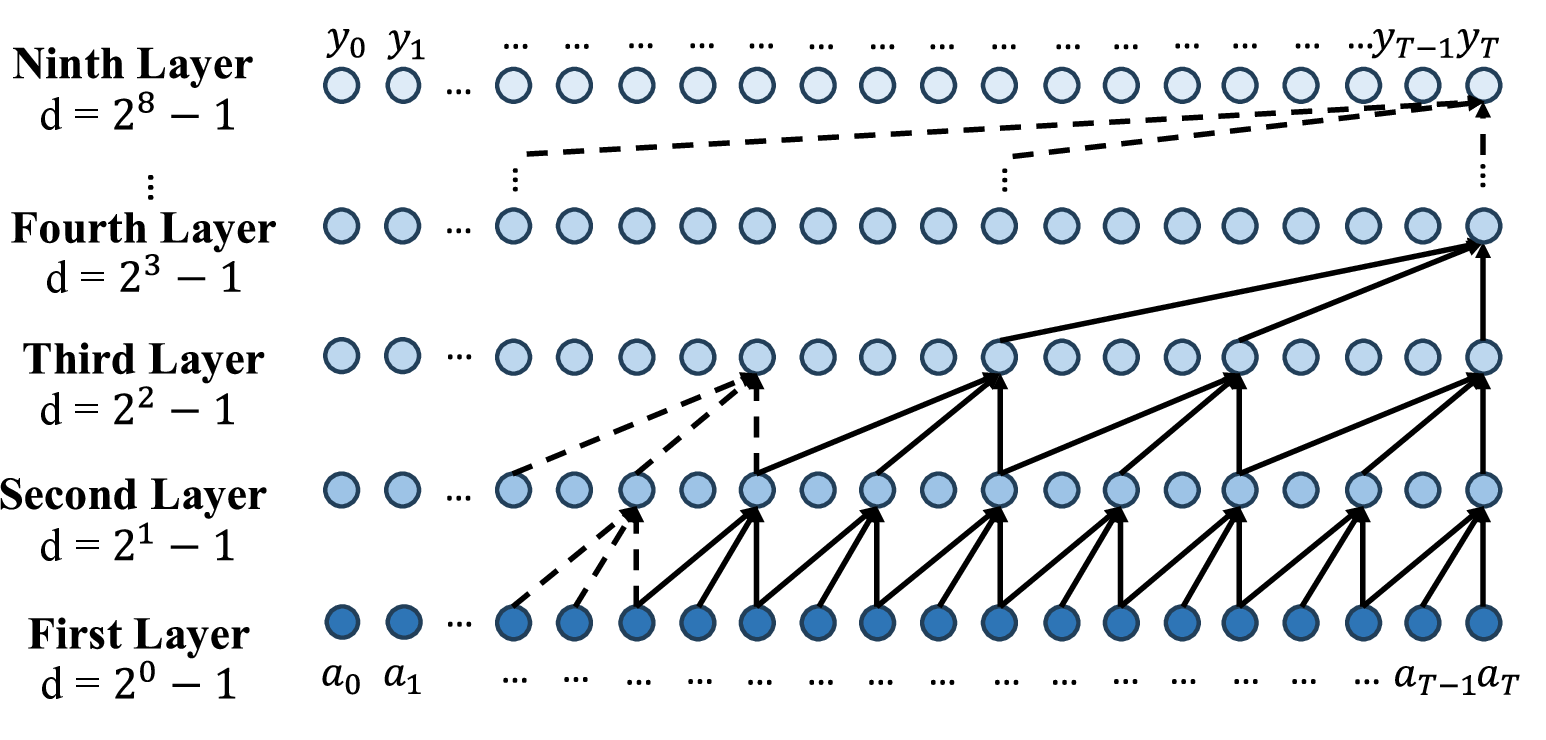}
    \caption{The structure of the $9$-layer TCN with dilation factor $D=2$ and kernel size $K=3$.}
    \label{fig:tcn_net} 
\end{figure}
\subsubsection{Long-Term ECG Reconstruction}
The long-term ECG reconstruction network adopts a similar encoder-decoder-fusion structure with SCEG as shown in Figure~\ref{fig:radarODE}(c), with the detailed structure and parameters shown in Table~\ref{tab:long_structure} and Figure~\ref{fig:tcn_net}. The encoder takes $50$ series of 4-sec time-domain radar signals as input and is composed of three groups of residual and downsample blocks with the deformable 2D convolution in Figure~\ref{fig:backbone} replaced by 1D convolution. Then, the decoder is realized by three Transconv1d blocks to generate the temporal feature.

The temporal feature generated from decoder in Table~\ref{tab:long_structure} will be stacked with morphological feature as illustrated in Figure~\ref{fig:radarODE}(c), and these two features act as two channels for later feature fusion by the temporal convolutional network (TCN). In general, TCN adopts dilation 1D convolution (Dilated Conv1d) to process multi-channel data structure using expanded receptive field with gaps between elements as shown in Figure~\ref{fig:tcn_net}, and the feature fusion is achieved during channel reduction as a common technique in traditional convolution neural network~\cite{xue2023novel}. Specifically, the solid line in Figure~\ref{fig:tcn_net} shows the connection of a $4$-layer TCN with the gaps for each layer $l$ as $d=D^{l-1}-1$, and the output $y_T$ is predicted based on a receptive field of $15$ input feature points $\{a_{T-14},\dots,a_T\}$. In this paper, $9$-layer Dilated Conv1d is adopted with dilation factor $D=2$ and kernel size $K=3$, and the receptive field is $511$ to make the most of contextual information contained in the temporal and morphological features.

\section{Dataset and Implementation Details}\label{sec:dataset}
\subsection{Hardware and Environment Settings for Data Collection}
The public dataset can be requested from~\cite{chen2022contactless} and is collected by the TI AWR-1843 radar with $77$~GHz start frequency and $3.8$~GHz bandwidth, providing good SNR and resolution to detect subtle vibrations that fit the proposed signal model and framework. To realize the 3D beamforming to extract cardiac features from real 3D space, $3$ transmitters (Tx) and $4$ receivers (Rx) are enabled with time division multiplexing multi-input multi-output (TDM-MIMO) applied in chirp transmitting and receiving, and the essential parameters in radar configuration setting are listed in Table~\ref{tab:data_param}.

\begin{table}[tb]
\centering
\caption{Parameters for Data Collection Interface}
    \begin{tabular}{lc?lc}
    \toprule
    \textbf{Parameter} & \textbf{Value} & \textbf{Parameter} & \textbf{Value} \\
    \toprule
    Start Frequency & $77$ GHz & Frequency Slope & $65$ MHz/$\mu$s \\ 
    Idle Time & $10$ $\mu$s    & Tx Start Time & $1$ $\mu$s \\
    ADC Start Time & $6$ $\mu$s  & ADC Samples & $200$ \\ 
    Sample Rate & $5000$ kbps  & Ramp End Time & $60$ $\mu$s \\ 
    Rx Gain & $30$ dB  & Rx Gain Target & $30$ dB  \\ 
    Start/End Chirp Tx & $0/2$  & No. of Chirp Loops & $2$  \\ 
    No. of Frames & $3600$  & Frame Periodicity & $50$ ms  \\ 
    \bottomrule
    \end{tabular}
\label{tab:data_param}
\end{table}%

\begin{figure}[tb] 
    \centering 
    \includegraphics[width=0.9\columnwidth]{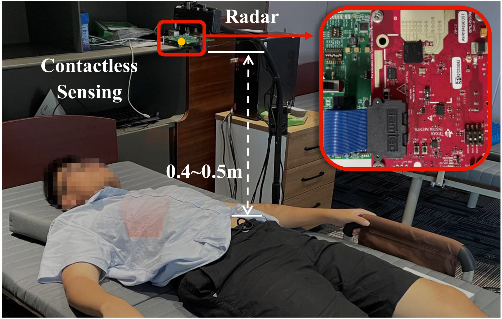}
    \caption{Environment settings for data collection from quasi-static subject~\cite{chen2022contactless}.}
    \label{fig:env_set} 
\end{figure}

The data collection is performed for subjects lying on the bed with quasi-static status to ensure good SNR with the least RBM noise. In addition, radar is placed right above the human chest region in a range of $0.4-0.5$m with minor propagation attenuation as shown in Figure~\ref{fig:env_set}, and hence the large- or small-scale signal variations (e.g., path loss, multi-path fading) are not considered in~\cite{chen2022contactless}.

\subsection{Link Budget Analysis}
Link budget analysis is a common evaluation for the performance of a radar system by accounting for all gains and losses from the transmitter to the receiver~\cite{mercuri2019vital,lin2022broadband}, but such analysis is not provided in~\cite{chen2022contactless}. For a radar system, the received power $P_{\mathrm{R}}$ can be expressed with respect to the transmitted power $P_{\mathrm{T}}$ as:
\begin{equation}\label{equ:proga}
P_{\mathrm{R}}=\frac{G_{\mathrm{T}} G_{\mathrm{R}} \lambda^2 \sigma}{(4 \pi)^3 R^4} P_{\mathrm{T}}
\end{equation}
where $G_T$/$G_R$ is the gain of the transmitter/receiver, $\sigma$ is the effective radar cross section (RCS) for the human chest region, and $R$ is the distance between radar and chest~\cite{mercuri2019vital}.

In addition, the lowest detectable signal power of the receiver is defined in~\cite{lin2022broadband} as:
\begin{equation}\label{equ:p_desire}
P_{\mathrm{r}, \min}= -174\ \mathrm{dBm} + 10\log_{10}(B) + \mathrm{NF} + \mathrm{SNR}_{\min}
\end{equation}
where $-174$ represents the thermal-noise level, NF means the noise figure, $B$ is the bandwidth of the receiver and $\mathrm{SNR}_{\min}$ is the desired SNR considering the afterward signal processing.

By combining (\ref{equ:proga}) and (\ref{equ:p_desire}), the maximum detectable range can be calculated as:
\begin{equation}\label{equ:r_max}
R_{\text {max}}=\sqrt[4]{\frac{G_{\mathrm{T}} G_{\mathrm{R}} \lambda^2 \sigma P_{\mathrm{T}}}{(4 \pi)^3 \ P_{\mathrm{r}, \min}}}
\end{equation}

All the radar-related values in (\ref{equ:r_max}) are shown in Table~\ref{tab:link_param} according to the datasheet of TI AWR-1843 radar~\cite{AWR1843}. According to the previous work~\cite{lin2022broadband,chen2022contactless}, the RCS value can be estimated as $-20$ dBsm for the quasi-static subjects wearing electrically thin cloth (i.e., the thickness is much smaller than the wavelength), with respiration noise filtered during the pre-processing stage. After substituting the values, the lowest detectable signal power is $P_{\mathrm{r}, \min}=-53$ dBm, and the maximum detectable range is $R_{\text {max}}=4$ m, revealing that the parameter setting used in~\cite{chen2022contactless} could provide the received signal with good SNR.

\begin{table}[tb]
\centering
\caption{Parameters for Link Budget Analysis}
    \begin{tabular}{lc?lc}
    \toprule
    \textbf{Parameter} & \textbf{Value} & \textbf{Parameter} & \textbf{Value} \\
    \toprule
    Tx Gain ($G_T$) & $10$ dBi  & Rx Gain ($G_R$) & $30$ dBi  \\
    Tx power ($P_{\mathrm{T}}$)& $12$ dBm & Noise Figure (NF)     & $15$ dB   \\ 
    Wavelength ($\lambda$)  & $3.9$ mm               & Bandwidth (B)         & $3.8$ GHz \\
    RCS ($\sigma$)     & $-20$ dBsm  &  Desired SNR ($\mathrm{SNR}_{\min}$) & $10$ dB  \\
    \bottomrule
    \end{tabular}%
\label{tab:link_param}
\end{table}%

\subsection{Dataset Description}
Half of the actual dataset is released with $91$ trials for $11$ subjects (with subject ID 1, 2, 5, 9, 10, 13, 14, 16, 17, 29, 30), and each trial contains $3$ minutes of data (radar measurements and ECG ground truth) with $200$~Hz sampling rate collected under $4$ physiological statuses (i.e., normal breath (NB, $43$ trials), irregular breath (IB, $18$ trials), sleep (SP, $18$ trials) and post exercise (PE, $12$ trials)). In addition, the work in~\cite{chen2022contactless} has pre-processed the radar signal using several techniques, such as 3D beamforming, dynamic time wrapping, to remove respiration noise and enhance cardiac activities. Lastly, no existing study is found for ECG reconstruction in literature based on the same dataset, and the proposed framework MMECG in~\cite{chen2022contactless} will be used as the only benchmark to make a comparison with our radarODE.

\subsection{Implementation Details and Compared Frameworks}
The proposed radarODE network is coded using PyTorch and trained for $200$ epochs with batch size $32$ on the NVIDIA RTX A4000 ($16$~GB) using stochastic gradient descent optimizer with early stop function~\cite{yao2007early} and learning rate $0.001$ based on a cosine annealing schedule~\cite{loshchilov2016sgdr}. The dataset is split into training and testing sets based on $11$-fold cross-validation with $1$ fixed subject for testing and the other $10$ subjects alternatively selected for training or validation, ensuring to make the most of all the trials while excluding the testing data from the training phase. In addition, all the ground truth characteristic peaks, PPI, and cardiac cycles are obtained by the NeuroKit2 from ECG signals~\cite{makowski2021neurokit2}. Furthermore, the $4$-seconds-long input SST segments only contain the frequency component within $[1,25]$~Hz and are down-sampled to $30$~Hz in the time-axis for saving memory usage in backbone design. This research has been approved by University Ethics Committee of Xi'anJiaotong-Liverpool University with proposal number ER-SAT-0010000090020220906151929.

In addition, three frameworks are selected for comparison with the following brief introduction of the architecture:
\begin{itemize}
  \item MMECG~\cite{chen2022contactless} receives multiple 1D radar signals as input and utilizes Conv1d and Transformer as decoder to simultaneously extract temporal and spatial features. The encoded features are further fused by multiplication and then decoded by Transconv1d and TCN to produce ECG recovery. 
  \item RSSRNet~\cite{wu2023contactless} takes spectrogram (STFT) as input with a Conv2d backbone and Transformer encoder. The adopted decoder is Transconv2d, and the output is still a spectrogram and needs to be converted to the ECG signal via inverse STFT.
  \item RadarNet~\cite{li2024radarnet} adopts 1D radar signal as input and directly generates coarse ECG signal using Conv1d. Then, several layers of ResNet are adopted to refine the ECG waveform.
\end{itemize}

\section{Experimental Results and Evaluations}\label{sec:exp}
This section provides the experimental results and evaluations in terms of three core modules as depicted in Figure~\ref{fig:radarODE}, with the first module providing PPI estimation for input/output slicing and reshaping, the second module robustly generating ECG pieces for the single cardiac cycle, and the third module yielding the final long-term ECG recovery.

\subsection{Evaluations of PPI Estimation} \label{sec:ppi}
\begin{figure}[tb]
        \centering
        \subfloat[]{\label{fig:ppi_obj}\includegraphics[width=1\columnwidth]{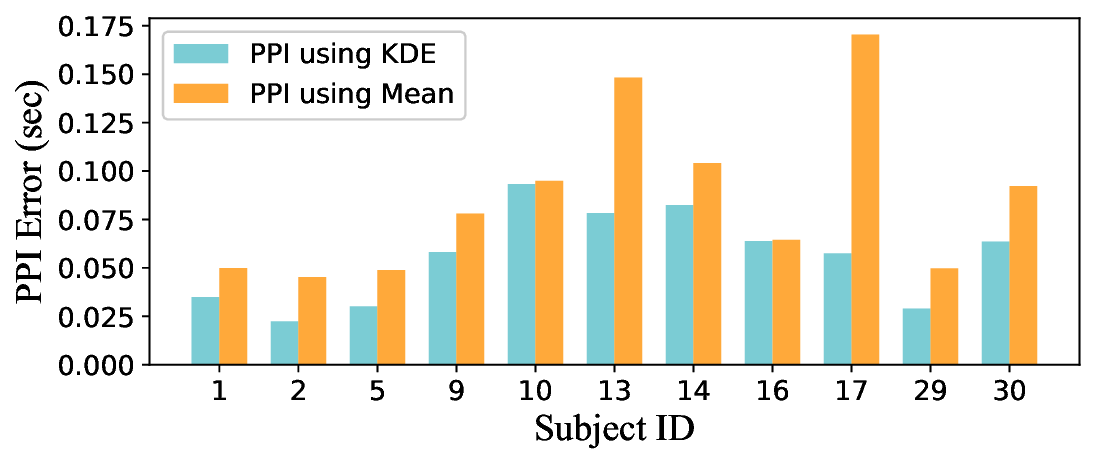}}\\ \vspace{-0.04\columnwidth}
        \subfloat[]{\label{fig:ppi_phy}\includegraphics[width=0.5\columnwidth]{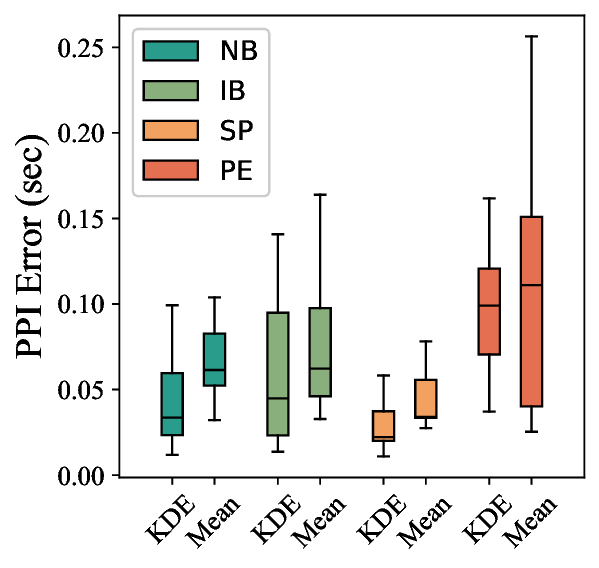}}
        \subfloat[]{\label{fig:ppi_cdf}\includegraphics[width=0.49\columnwidth]{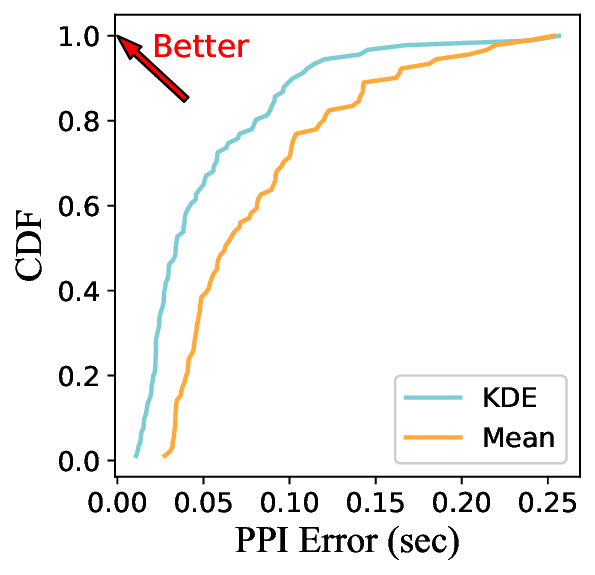}}
        \caption{PPI estimation error of KDE-based and mean-based method: (a) View of different subjects; (b) View of different physiological statuses; (c) CDF for the overall PPI estimation error.}
        \label{fig:ppi_compare}
\end{figure}

The PPI estimation is the first module of radarODE, and the accuracy of the estimated PPI directly affects the fidelity of the concatenated morphological reference. Therefore, Figure~\subref*{fig:ppi_obj} shows the PPI estimation error obtained by Algorithm~\ref{alg:ppi} for all subjects based on KDE defined in (\ref{equ:kde}) or directly averaging the candidate PPI values. The large PPI error (e.g., for subject $13$, $17$) is normally caused by body movements or residual respiration noise due to IB or PE status, as also shown in Figure~\subref*{fig:ppi_phy} with large median error and variation for both methods. In contrast, the subjects in NB and SP statuses tend to be stable with less body movement, and the respiration noise can be well eliminated, achieving low median PPI errors as $0.03$s and $0.02$s using the KDE-based method for each status.

Overall, it is clear that the KDE-based PPI estimation is more accurate than the mean-based estimation for each subject, as shown in the cumulative distribution function (CDF) in Figure~\subref*{fig:ppi_cdf}, because the KDE-based method is robust to the outliers caused by noises and could figure out the correct PPI estimation near the majority of candidate values.

\begin{figure}
  \centering
  \subfloat[]{\label{fig:rmse_cdf_compare}\includegraphics[width=0.5\columnwidth]{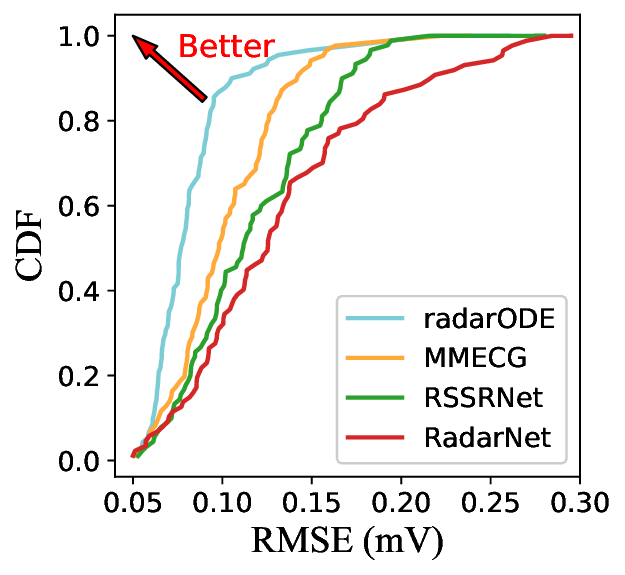}}
  \subfloat[]{\label{fig:sceg_cor_cdf}\includegraphics[width=0.48\columnwidth]{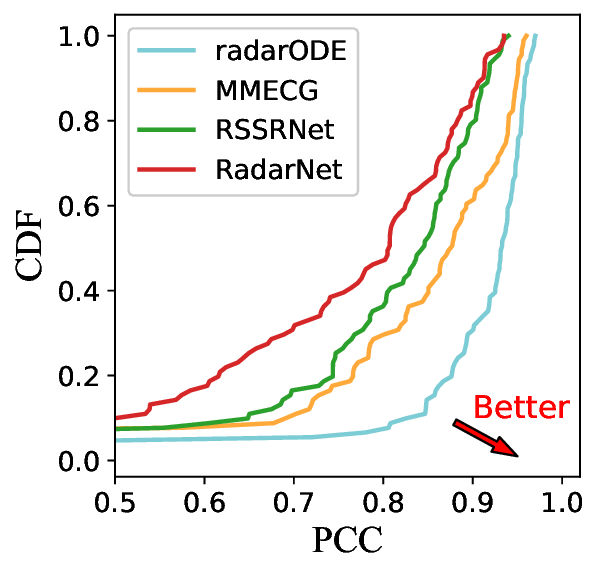}}\\
  \subfloat[]{\label{fig:MDR_cdf_overall}\includegraphics[width=0.49\columnwidth]{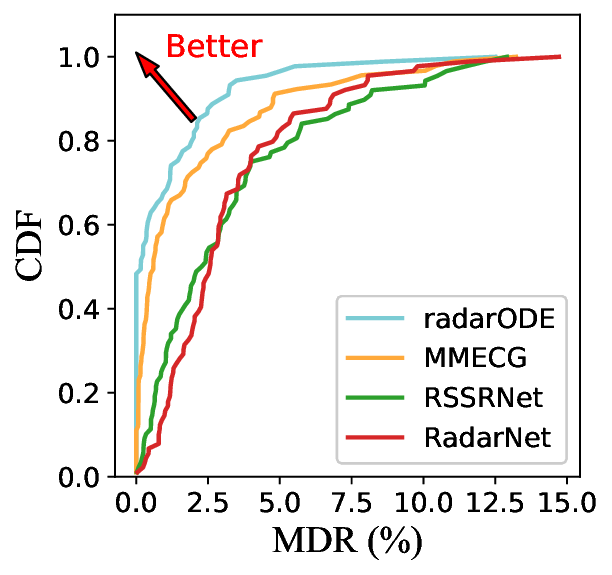}}
  \subfloat[]{\label{fig:sceg_r_cdf}\includegraphics[width=0.48\columnwidth]{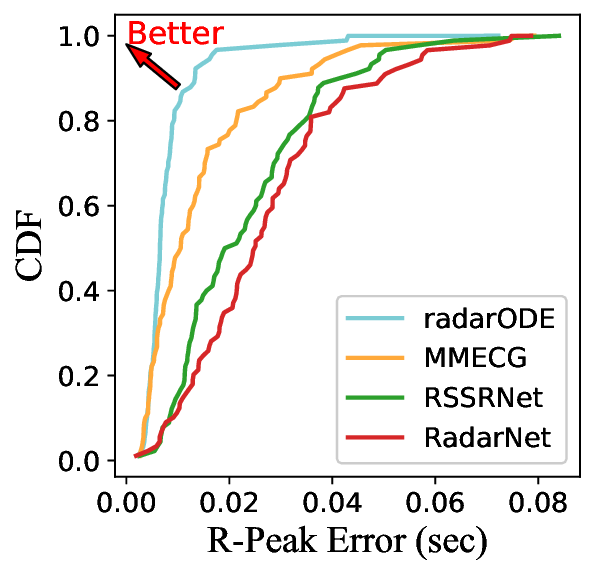}}
  \caption{Performance comparison for single-cycle ECG recovery: (a) - (d) CDF of RMSE, PCC, MDR and R-peak error for all trials.}
  \label{fig:sceg_performance}
\end{figure}
\subsection{Evaluations of SCEG Module in radarODE}
SCEG is the core module that realizes the domain transformation and ensures the robust long-term ECG recovery in the next stage, and the performance is evaluated on the generated single-cycle ECG pieces in terms of morphological accuracy, corrupt ECG reconstruction and absolute R-peak error as shown in Figure~\ref{fig:sceg_performance} and Table~\ref{tab:ablation}.

\subsubsection{Comparison of Morphological Accuracy}
The morphological accuracy is shown in Table~\ref{tab:ablation} as the median value of root mean square error (RMSE) and Pearson-correlation coefficient (PCC), with RMSE sensitive to the deviation of peaks and PCC focusing on the general shape. The overall performances for all trials are shown as CDF plots in Figure~\subref*{fig:rmse_cdf_compare} and~\subref*{fig:sceg_cor_cdf}. The results indicate that radarODE could generate high-fidelity ECG signals with good RMSE and PCC owing to the prior knowledge provided by the ODE decoder, while MMECG achieves the second-best result and shows a domain transformation ability for the majority of radar inputs with good SNR. In contrast, RSSRNet and RadarNet achieve a similar performance because they only accept single-channel inputs and cannot effectively leverage the features of $50$ channels provided in the dataset.

\subsubsection{Comparison of Corrupt ECG Recovery}
The ability to recover ECG pieces from corrupt radar signals reveals the noise robustness of each framework, and missed detection rate (MDR) is adopted to count failed recoveries without showing characteristic ECG patterns (R peaks). Three rules are made to define the failed ECG recoveries:
\begin{itemize}
\item The deviation of the corrupt R peak from the ground truth exceeds the absolute tolerance of $0.15$s~\cite{chen2022contactless}.
\item The corrupt R peak has one or more neighboring R peaks closer than $0.3$s.
\item The amplitude of the corrupt R peak is $30\%$ lower than that of the ground truth R peak.
\end{itemize}

The results of MDR are shown in Table~\ref{tab:ablation} with CDF plots of all trials shown in Figure~\subref*{fig:MDR_cdf_overall}. It is evident that radarODE achieves better noise robustness compared with previous work, owing to the constraint brought by ODE decoder. MMECG still has a better performance compared with RSSRNet and RadarNet due to the powerful backbone to extract features from undistorted channels. The overall performance of MDR in Figure~\subref*{fig:MDR_cdf_overall} coincides with the morphological accuracy in Figure~\subref*{fig:rmse_cdf_compare} and~\subref*{fig:sceg_cor_cdf}, because the corrupt ECG recoveries also significantly affect the RMSE and PCC.

\subsubsection{Comparison of R-peak Timing Error}
After filtering the corrupt ECG pieces, the absolute timing error of R peaks is calculated to evaluate the quality of fine-grained ECG features, with the median value and CDF plots shown in Table~\ref{tab:ablation} and Figure~\subref*{fig:sceg_r_cdf} respectively. The results are still similar to previous evaluations, with radarODE recovering the most accurate R peaks and the other three frameworks showing larger R-peak deviation. In addition to the benefits brought by ODE decoder, the adopted SST inputs also provide necessary time-frequency features to help the deep learning model identify indistinctive vibrations under low-SNR scenarios.

\begin{table*}[tb]
\centering
\caption{Performance Comparison for Single-cycle ECG Recovery and Ablation Study}
    \begin{tabular}{c c c c |cccc}
    \toprule
    Framework & Backbone & Encoder & Decoder  & RMSE (mV) & PCC & MDR & R Error (sec)\\
    \toprule
    MMECG~\cite{chen2022contactless} & - & Conv1d + Transformer & Transconv1d + TCN & $0.091$  & $87.9\%$ & $1.24\%$ & $0.012$ \\
    RSSRNet~\cite{wu2023contactless} & Conv2d & Transformer & Transconv2d  & $0.100$ & $86.0\%$ & $2.17\%$ & $0.019$\\
    RadarNet~\cite{li2024radarnet} & - & Conv1d & Conv1d (ResNet) & $0.113$ & $80.2\%$ & $2.58\%$ & $0.024$\\
    \hline
    \multirow{3}*{radarODE}  & \multirow{3}*{\makecell[c]{Deform\\Conv2d}} & \multirow{3}*{Conv1d}  & Initial + Temporal & $0.086$  & $89.4\%$  & $1.53\%$ & $0.012$\\
    & &  & Initial + ODE & $0.092$ & $85.5\%$ & $\mathbf{0.14\%}$ & $\mathbf{0.005}$ \\
    & &  & Initial + Temporal + ODE  & $\mathbf{0.077}$ & $\mathbf{92.6\%}$ & $0.18\%$ & $0.006$\\
    \bottomrule
    \end{tabular}%
\label{tab:ablation}
\end{table*}%

\begin{figure}
  \centering
  \subfloat[]{\label{fig:ode_sceg}\includegraphics[width=0.5\columnwidth]{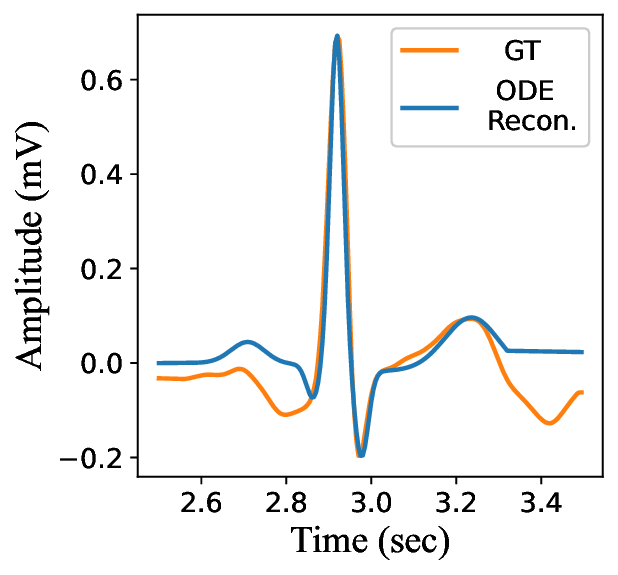}}
  \subfloat[]{\label{fig:loss_compare}\includegraphics[width=0.5\columnwidth]{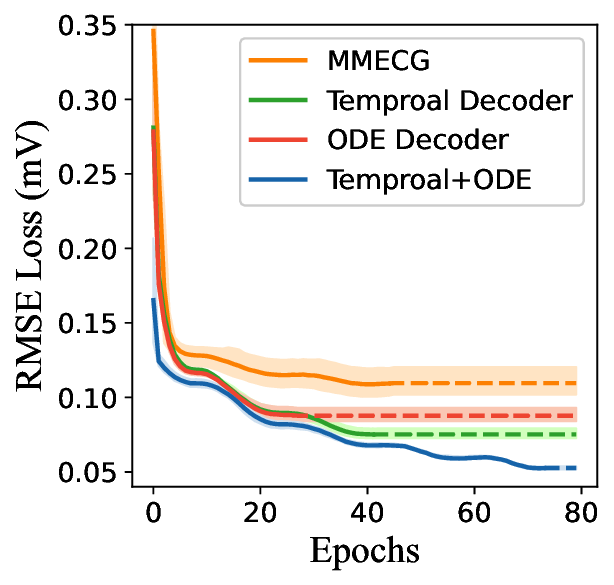}} \\ \vspace{-0.04\columnwidth}
  \subfloat[]{\label{fig:sceg_good}\includegraphics[width=0.48\columnwidth]{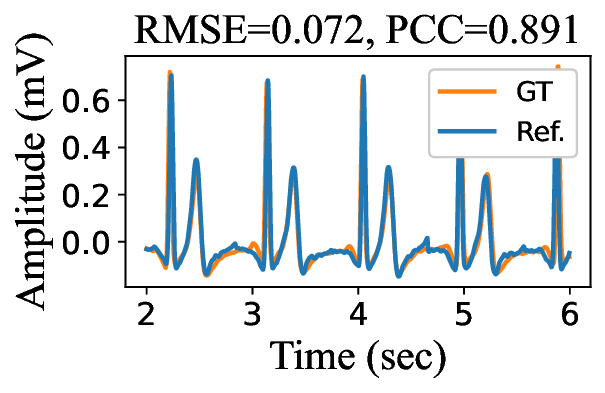}}
  \subfloat[]{\label{fig:sceg_bad}\includegraphics[width=0.5\columnwidth]{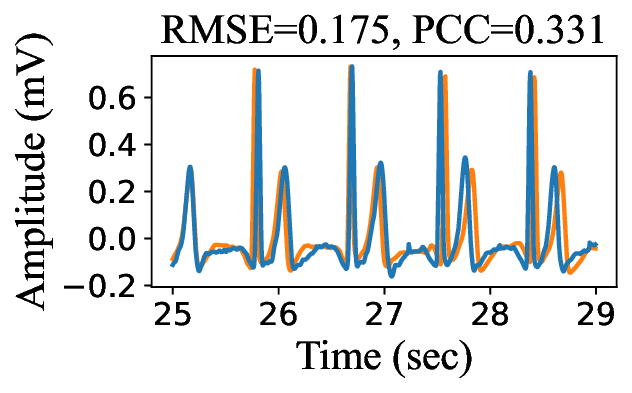}}
  \caption{Ablation study and visualization: (a) Rigid reconstruction (ODE Recon.) from ODE decoder with ground truth (GT); (b) Training loss comparison; (c) and (d) High- and Low-fidelity morphological references (Ref.) caused by PPI estimation error.}
  \label{fig:sceg_ablation}
\end{figure}
\subsubsection{Ablation Study} 
The ablation study is performed to further evaluate the contributions of temporal and ODE decoders. The results in Table~\ref{tab:ablation} reveal that both decoders can work individually and achieve reasonable results, but the ODE decoder has lower accuracy because it neglects the subtle ECG feature while only focusing on the characteristic peaks with rigid connections elsewhere, as shown in Figure~\subref*{fig:ode_sceg}. On the contrary, the introduced ODE model could resist strong noise and achieve the lowest MDR as $0.14\%$, while the temporal decoder only gets a similar MDR ($1.53\%$) with MMECG ($1.24\%$). In radarODE, the outputs of temporal and ODE decoder are fused together to achieve noise robustness with MDR$=0.18\%$, while maintaining a faithful ECG shape (i.e., good RMSE and PCC).

\begin{figure*}[tbp]
        \centering
    \begin{minipage}[t]{0.8\columnwidth}
    \subfloat[]{\label{fig:overall_ode}\includegraphics[width=1\columnwidth,valign=t]{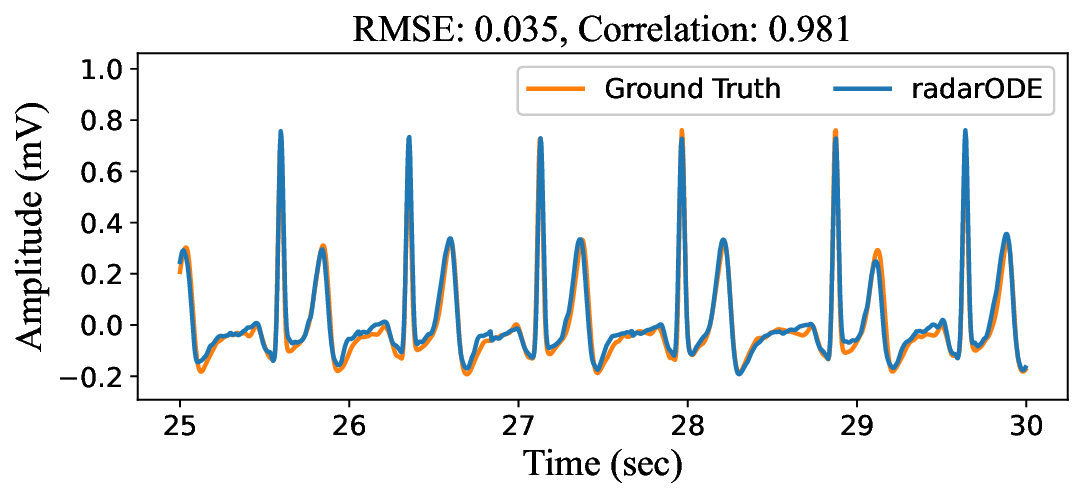}} \\  
    \subfloat[]{\label{fig:overall_mmecg}\includegraphics[width=1\columnwidth]{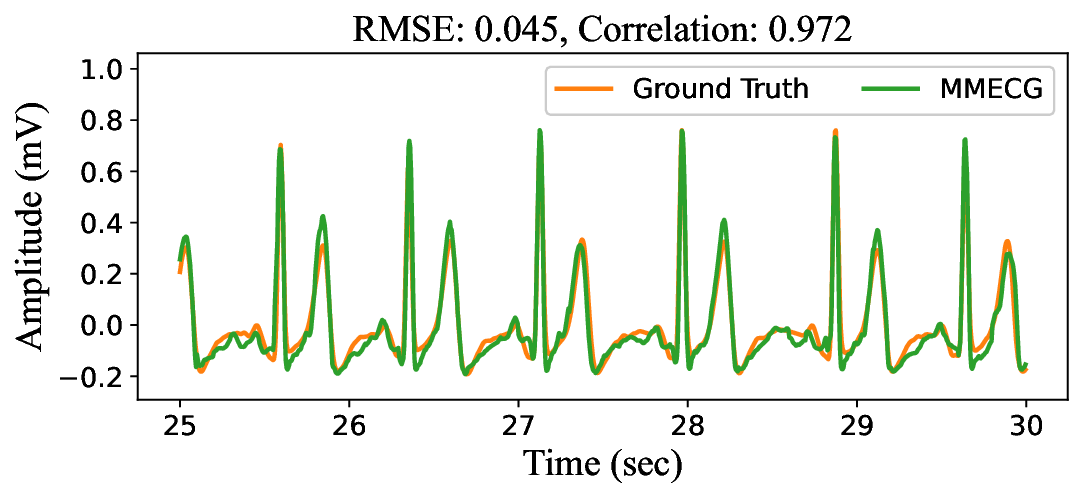}}
    \end{minipage}%
    \begin{minipage}[t]{0.8\columnwidth}
    \subfloat[]{\label{fig:overall_noise}\includegraphics[width=1\columnwidth,valign=t]{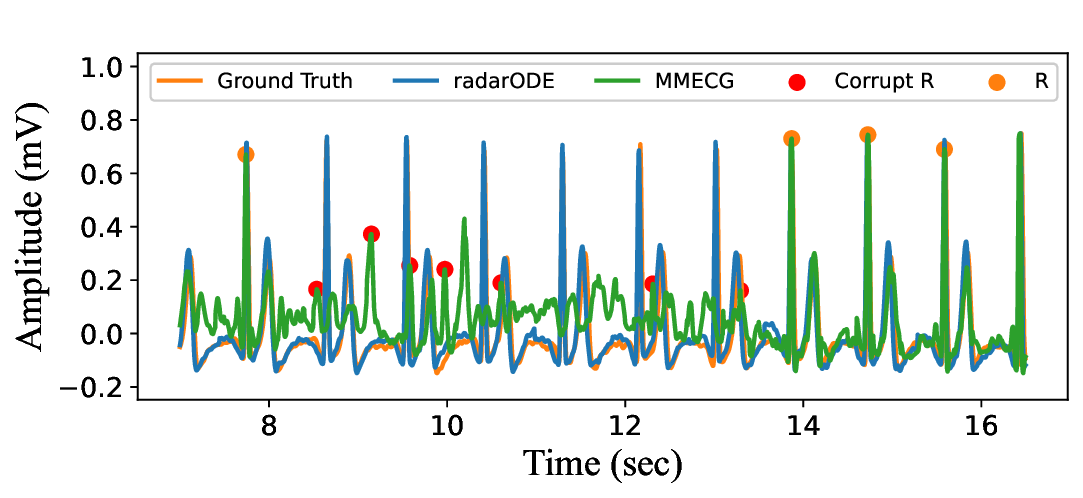}} \\
    \subfloat[]{\label{fig:overall_rcg}\includegraphics[width=1\columnwidth]{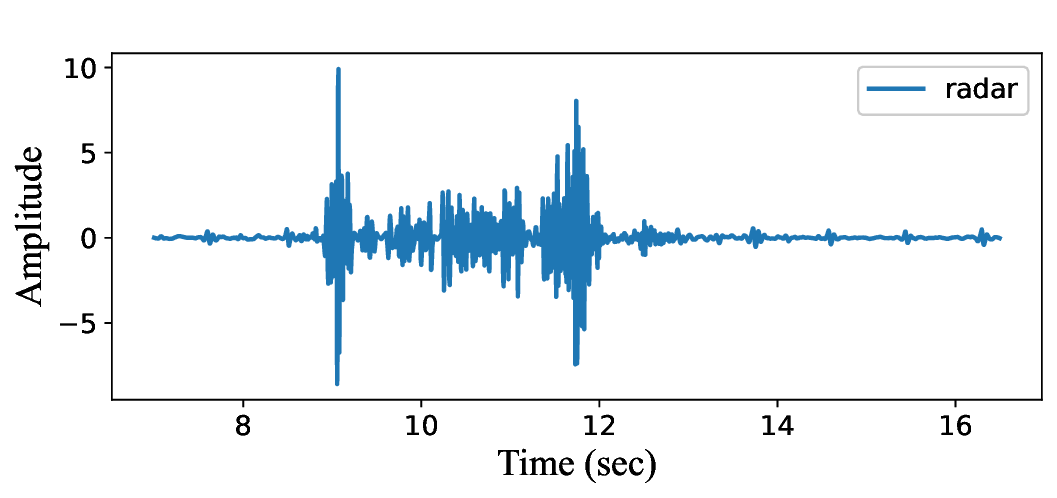}}
    \end{minipage}%
    \begin{minipage}[t]{0.4\columnwidth}
    \subfloat[]{\label{fig:detection_cdf_overall}\includegraphics[width=1\columnwidth]{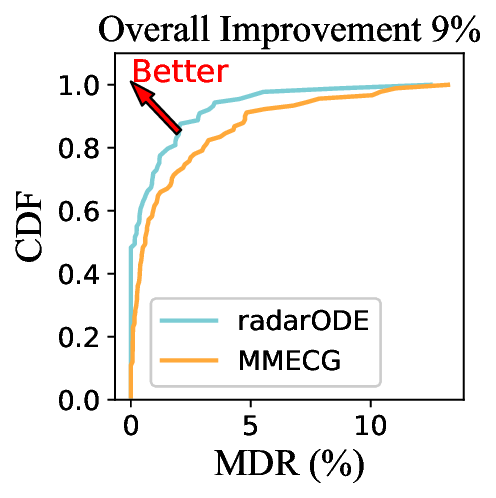}} \\
    \subfloat[]{\label{fig:detection_cdf_phy}\includegraphics[width=1\columnwidth]{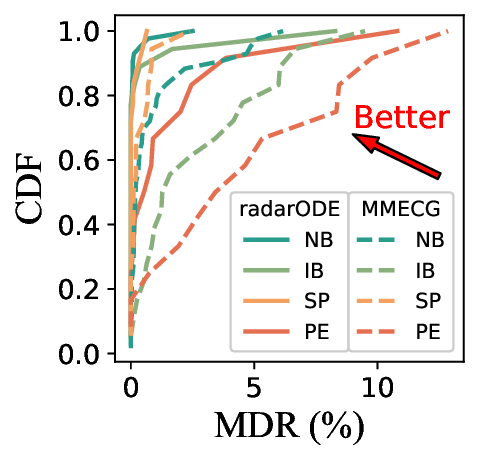}}
    \end{minipage}%
    \caption{Corrupt ECG reconstruction: (a) and (b) Ideal reconstruction results for radarODE and MMECG; (c) Corrupt ECG reconstruction yield by MMECG and faithful reconstruction from radarODE; (d) Corresponding radar signal during body movements; (e) Overall missed detection rate (MDR); (f) MDR under different physical statuses.}
    \label{fig:overall_compare}
\end{figure*}

\subsubsection{Visualization of Training Process} 
Figure~\subref*{fig:loss_compare} illustrates the training loss for benchmark and ablation study with the dotted line representing the early stop of the training process and the shaded area indicating the repetition of the training process for five times. From the ablation study, SCEG with only ODE decoder cannot provide a very accurate reconstruction due to the rigid shape as shown in Figure~\subref*{fig:ode_sceg}, while the temporal decode could achieve the second-best result. After combining ODE and temporal decoders, the outputs from ODE decoder act as the morphological-prior to accelerate the convergence, achieving the RMSE of $0.17$mV after the first epoch. In addition, the morphological reference will not be destroyed by noises and could stabilize the training process, contributing to the lowest training loss as shown in Figure~\subref*{fig:loss_compare}.

\subsubsection{Summary of SCEG Evaluation} 
The proposed SCEG module in radarODE achieves better accuracy in generating single-cycle ECG pieces with the improvements brought by the ODE decoder and SST inputs, enabling successful recoveries even under abrupt noises and achieving the best RMSE, PCC and R-peak accuracy compared with previous frameworks. 

The generated ECG pieces can be resized and concatenated based on PPI estimation, but the PPI estimation error may accumulate and degrade the accuracy as shown in Figure~\subref*{fig:sceg_good} and~\subref*{fig:sceg_bad}, because slight deviations of the peaks ruin the overall RMSE/PCC, hence requiring long-term reconstruction module to refine the concatenated ECG pieces in the next step. To provide a comprehensive evaluation for long-term ECG recovery from various dimensions (e.g., trials, subjects, and physical statuses), MMECG is selected as the only benchmark in the next section for a clear result visualization.

\subsection{Overall Evaluations of radarODE}\label{sec:overall_ode}
The long-term ECG reconstruction module finally generates 3-minute-long ECG signals for the evaluations of the entire radarODE framework in terms of corrupt ECG reconstruction, morphological accuracy, and fine-grained cardiac feature accuracy.

\subsubsection{Corrupt ECG Reconstruction} 
The ideal reconstructed ECG signals are shown in Figure~\subref*{fig:overall_ode} and~\subref*{fig:overall_mmecg} with corresponding RMSE/PCC labeled. However, the long-term ECG signal may contain corrupt parts due to the presence of body movements (especially in IB and PE). Figure~\subref*{fig:overall_noise} shows the corrupted ECG reconstruction yield by MMECG with the falsely detected R peaks noted as red dots, and Figure~\subref*{fig:overall_rcg} is the raw radar signal with extensive distortion induced by body movements. In contrast, the radarODE could still provide faithful ECG reconstructions under body movements due to the introduction of prior knowledge (ODE model) about ECG as the morphological reference, hence gaining certain robustness in resisting noises.
\begin{figure*}[t]
        \centering
    \begin{minipage}[t]{0.5\columnwidth}
    \subfloat[]{\label{fig:rmse_cdf}\includegraphics[width=0.845\columnwidth,valign=t]{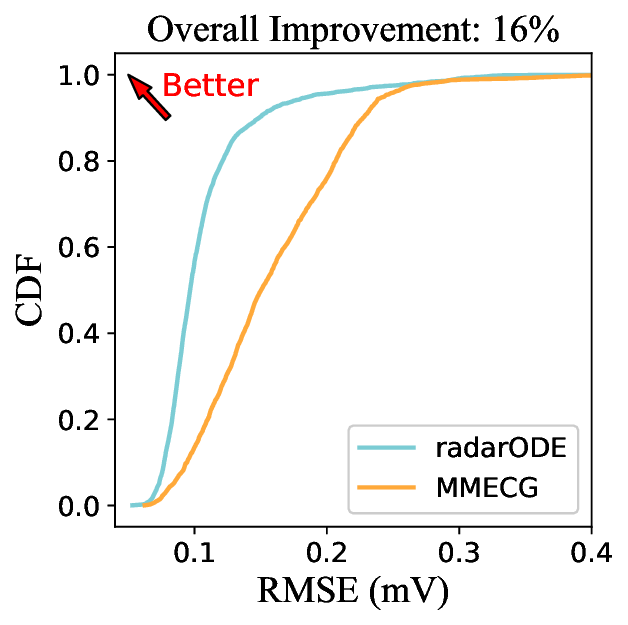}} \vspace*{0.02\columnwidth} \\ 
    \subfloat[]{\label{fig:cor_cdf}\includegraphics[width=0.845\columnwidth]{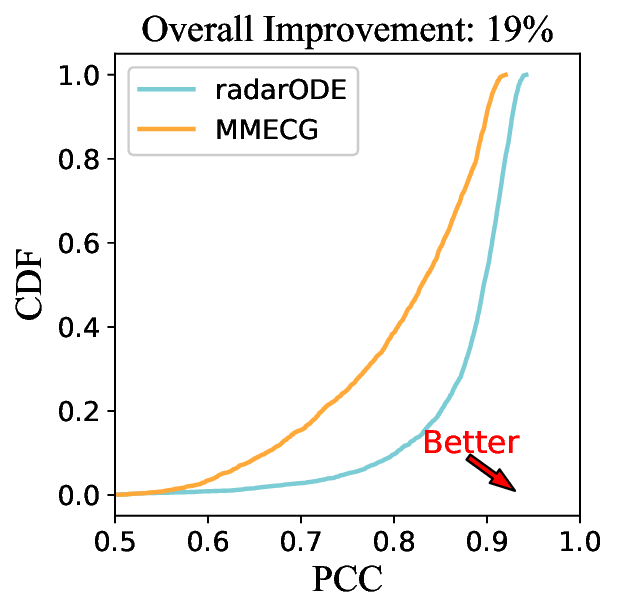}}
    \end{minipage}%
    \hspace{-0.1\columnwidth}
    \begin{minipage}[t]{1\columnwidth}
    \subfloat[]{\label{fig:rmse_obj}\includegraphics[width=1\columnwidth,valign=t]{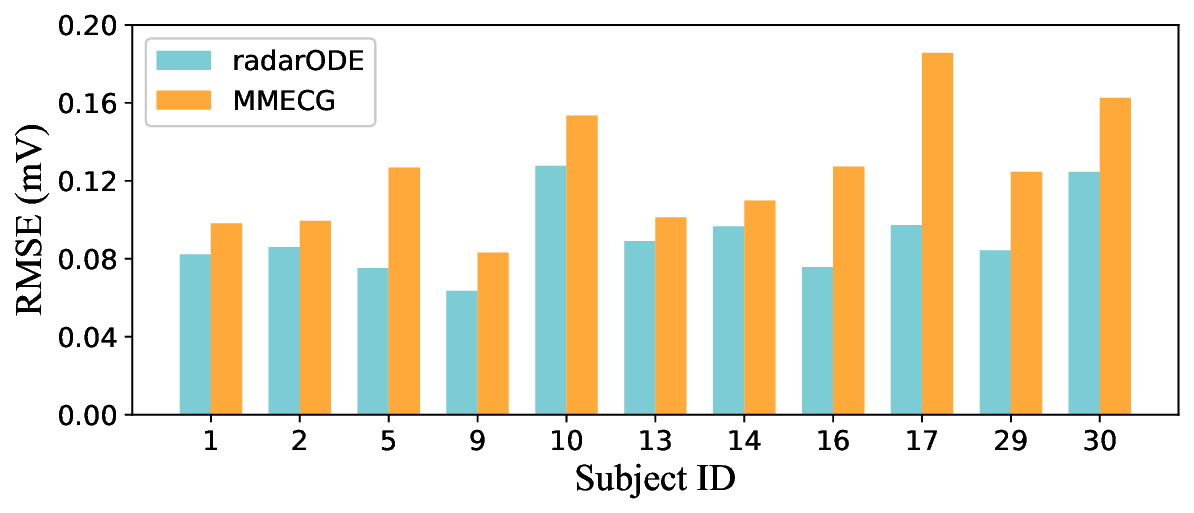}} \\
    \subfloat[]{\label{fig:cor_obj}\includegraphics[width=1\columnwidth]{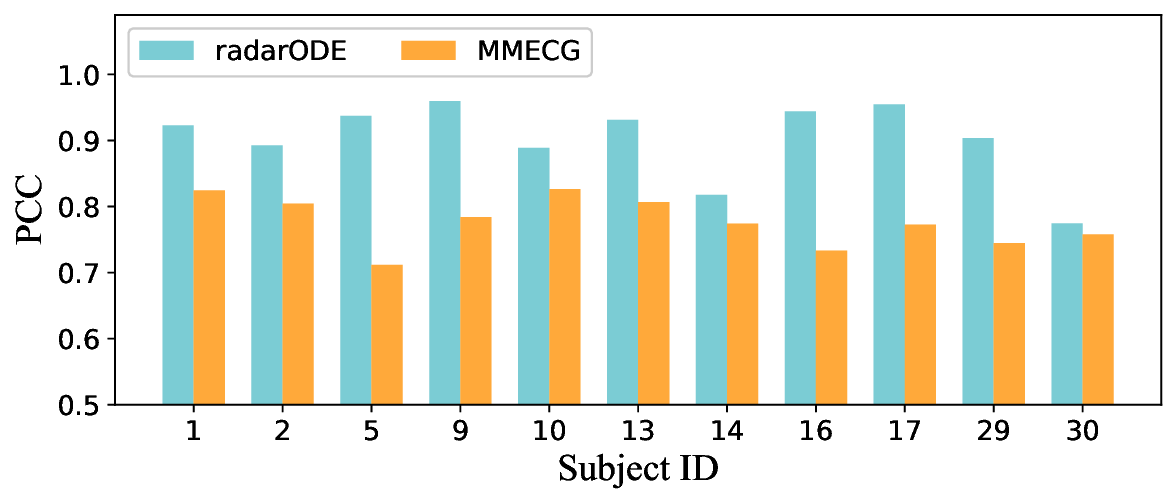}}
    \end{minipage}%
    \begin{minipage}[t]{0.5\columnwidth}
    \subfloat[]{\label{fig:rmse_phy}\includegraphics[width=0.8\columnwidth]{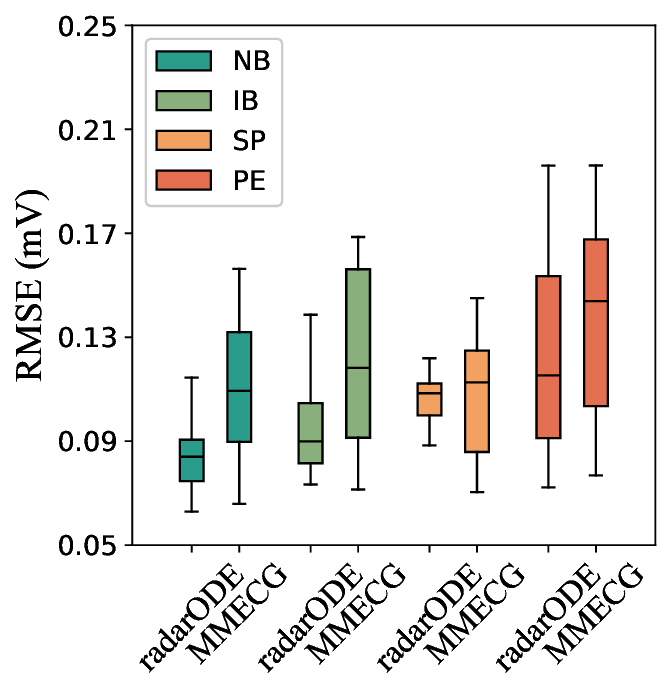}} \vspace*{0.05\columnwidth}\\
    \subfloat[]{\label{fig:cor_phy}\includegraphics[width=0.8\columnwidth]{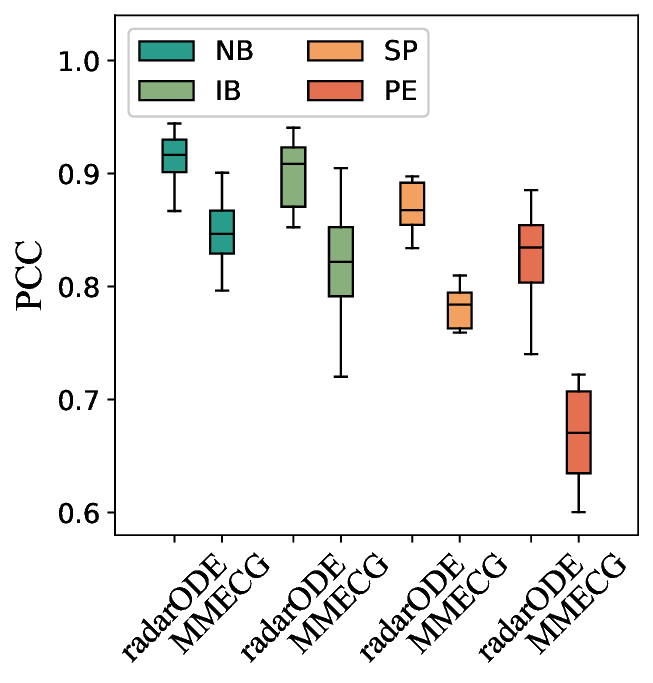}}
    \end{minipage}%
    \caption{Morphological accuracy comparison: (a) and (b) Overall CDF of RMSE and PCC for all trials; (c) and (d) RMSE and PCC under different physical statuses; (e) and (f) RMSE and PCC across all subjects.}
    \label{fig:rmse_cor}
\end{figure*}

The overall MDR can be calculated from the recovered ECG signal with the CDF plot shown in Figure~\subref*{fig:detection_cdf_overall}, and the overall improvement achieved by radarODE is $9\%$. In addition, the CDF plots for different physical statuses are plotted in Figure~\subref*{fig:detection_cdf_phy} with the $90$-percentile MDR of $0.12\%$, $0.85\%$, $0.32\%$, $3.71\%$ during NB, IB, SP, PE for radarODE and $3.36\%$, $6.22\%$, $0.83\%$, $9.64\%$ for MMECG. The result shows that different physical statuses have noticeable impacts on the quality of the reconstructed long-term ECG, and radarODE could provide a lower MDR than MMECG in all statuses owing to the prior knowledge in the ODE decoder.

\subsubsection{Morphological Accuracy}
The morphological accuracy measures the similarity between reconstructed and ground truth ECG signals by calculating RMSE and PCC. Figure~\subref*{fig:rmse_cdf} and~\subref*{fig:cor_cdf} show the overall performance of radarODE and MMECG in CDF with the median RMSE/PCC of $0.097$mV$/89.6\%$ and $0.120$mV$/81.2\%$. It is worth noticing that the overall improvement of radarODE compared to MMECG is $16\%$ and $19\%$ for RMSE and PCC across $91$ trials respectively, indicating that the morphological-prior is more helpful in generalizing the typical ECG pattern than calibrating the peaks.

In addition, Figure~\subref*{fig:rmse_obj} and~\subref*{fig:cor_obj} illustrate the RMSE/PCC across all subjects, with radarODE always achieving better results than MMECG. It is worth noting that the results of the long-term reconstruction show certain consistency with the previous PPI estimation error, because the fidelity of the morphological reference is directly affected by the PPI error. For example, subjects $10$, $13$, $14$, and $30$ get worse results than others in either RMSE or PCC evaluation due to the large PPI estimation error as shown in Figure~\subref*{fig:ppi_obj}.

Lastly, Figure~\subref*{fig:rmse_phy} and~\subref*{fig:cor_phy} illustrate the RMSE/PCC for all trials in terms of different physical statuses, and the box plots show that the stable statuses (i.e., NB, SP) guarantee the reconstruction with small variance. In contrast, unstable statues (i.e., IB, PE) can severely ruin the radar signal due to body movements, causing an inconsistent quality of the reconstructed ECG. However, radarODE could still provide the reconstructions with a smaller variance than MMECG, especially for unstable statues because of the morphological-prior embedded in the ODE decoder.

\subsubsection{Fine-Grained Cardiac Events Reconstruction}
\begin{figure}[!h]
  \centering
  \subfloat[]{\label{fig:peak_cdf}\includegraphics[width=1\columnwidth]{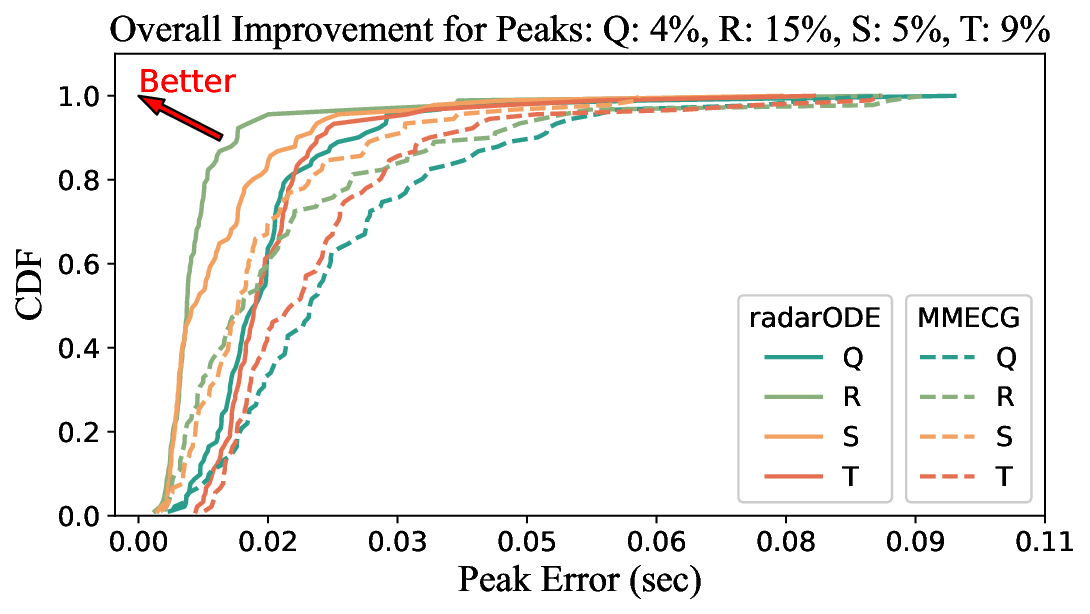}}\\ \vspace{-0.03\columnwidth}
  \subfloat[]{\label{fig:peak_phy_ode}\includegraphics[width=0.5\columnwidth]{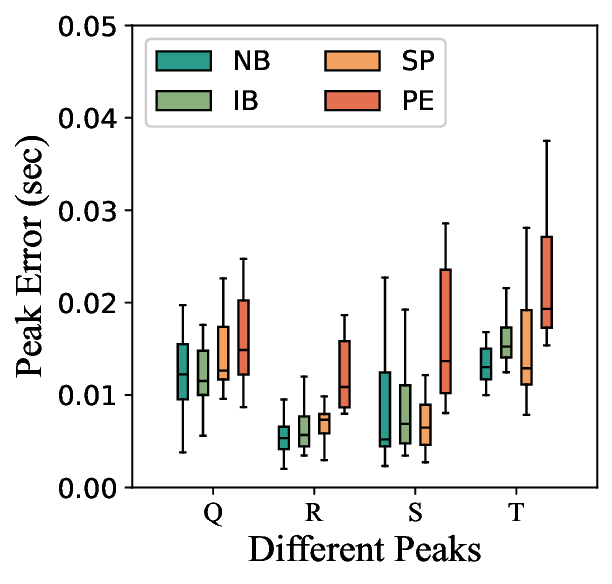}}
  \subfloat[]{\label{fig:peak_phy_mmecg}\includegraphics[width=0.5\columnwidth]{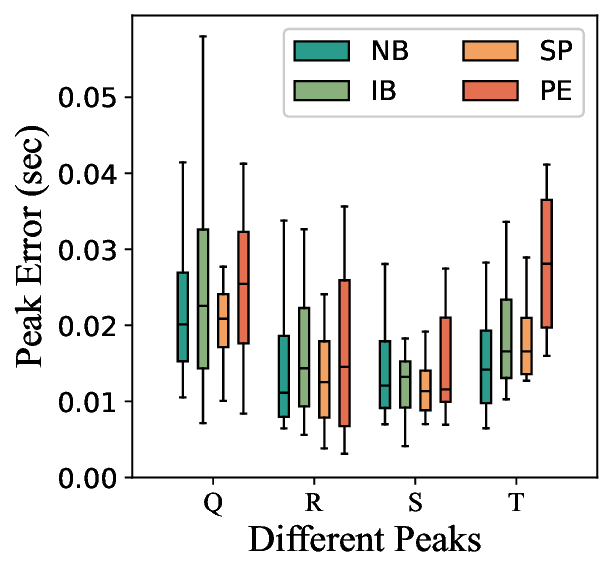}} 
  \caption{Evaluations of fine-grained cardiac events: (a) CDF of the overall timing error; (b) and (c) The timing error for different peaks under various physical statuses for radarODE and MMECG.}
  \label{fig:peak_compare}
\end{figure}

The evaluation of fine-grained cardiac features aims to analyze the timing accuracy of the QRST peaks and the P peak is not considered in this evaluation as also suggested in the benchmark paper~\cite{chen2022contactless}, because the P peak is inconspicuous and even unable to be detected for some ground truth ECG signal. The overall result is shown in Figure~\subref*{fig:peak_cdf} as CDF with the median/$90$-percentile absolute timing error for QRST peaks shown in Table~\ref{tab:peak_acc}. The improvement owes to the use of SST spectrogram with more evident patterns for the prominent heart vibrations, and the ODE decoder contributes to calibrating the peak positions according to $\eta$ and $\tau$, hence improving the overall peak accuracy.

\begin{table}[tb]
\centering
\caption{Absolute Timing Error for Reconstructed ECG Peaks}
    \begin{tabular}{c c cccc}
    \toprule
    Framework & Percentile & Q & R & S & T \\
    \toprule
    \multirow{2}*{MMECG~\cite{chen2022contactless}} & Median & $0.021$ & $0.012$ & $0.019$ & $0.018$ \\
                                           & $90$-percentile & $0.048$ & $0.023$ & $0.029$ & $0.035$ \\
    \toprule 
    \multirow{2}*{radarODE}                         & Median & $0.015$ & $0.007$ & $0.009$ & $0.014$ \\
                                           & $90$-percentile & $0.027$ & $0.015$ & $0.020$ & $0.023$ \\
    \bottomrule
    \multicolumn{6}{r}{unit: second} \\
    \end{tabular}%
\label{tab:peak_acc}
\end{table}%

Figure~\subref*{fig:peak_phy_ode} and~\subref*{fig:peak_phy_mmecg} demonstrate the improvement of radarODE in terms of different physical statuses, and the results coincide with many previous evaluations. Firstly, the radarODE outperforms MMECG for all statuses with the R peak always achieving the best accuracy. Secondly, stable physical statuses tend to yield accurate reconstruction with small variance, but the difference between statues is smaller than that of the morphological accuracy analysis, because the corrupt reconstructions have been filtered for peak accuracy evaluation.

\subsubsection{Summary of Long-term ECG Recovery Evaluation} 
The experimental results illustrated that the proposed radarODE could yield high-quality long-term ECG recovery even under extensive body movement noise, with better RMSE and PCC compared with the benchmark. In addition, the introduction of the SST spectrogram and the ODE decoder further improves the accuracy of the fine-grained cardiac features that are crucial to the potential applications in clinical diagnosis.

\begin{table*}[tb]
\centering
\caption{Complexity Analysis of Deep Learning Framework}
    \begin{tabular}{c|ccccc}
    \toprule
    {\multirow{2}*{\textbf{Framework}}} & \multicolumn{2}{c}{\qquad \qquad \qquad \textbf{Time Complexity}} & \textbf{Space Complexity} & \multirow{2}*{\textbf{Time/Epoch} (min)} \\
     & $\mathcal{O}(\cdot)$ per layer & FLOPs (G) & Params. (M) \\
    \midrule
    MMECG~\cite{chen2022contactless} & MHSA: $\mathcal{O}(CL^2)$, Conv1d: $\mathcal{O}(LKC^2)$  & $0.59$ & $0.67$ & $3.25$ \\
    RSSRNet~\cite{wu2023contactless} & MHSA: $\mathcal{O}(CL^2)$, Conv2d:$\mathcal{O}(HLK^2C^2)$ & $1.06$ & $4.51$ & $4.02$\\
    RadarNet~\cite{li2024radarnet} & Conv1d: $\mathcal{O}(LKC^2)$  & $0.24$ & $0.33$ & $2.16$\\
    \midrule
     radarODE & Conv2d: $\mathcal{O}(HLK^2C^2)$ & $1.45$  & $6.04$  & $4.51$ \\
    \bottomrule
    \multicolumn{5}{r}{MHSA: Multi-head self-attention, L: Length of input, H: Height of input, K: Kernel size, C: Channel}
    \end{tabular}
\label{tab:complex}%
\end{table*}%
\subsection{Complexity Analysis and Comparison}
Although the proposed radarODE performs better than other frameworks, it is necessary to analyze the complexity of the deep learning model for a fair comparison. In this subsection, the model complexity will be analyzed in terms of time and space complexity, and the training time for each epoch is also provided for an intuitive comparison as shown in Table~\ref{tab:complex}. 

\subsubsection{Time Complexity}
Time Complexity in deep learning typically refers to the computational cost during the model training and can be expressed in terms of floating-point operations (FLOPs) or Big O notation $\mathcal{O}(\cdot)$, with FLOPs measuring the floating-point calculations and $\mathcal{O}(\cdot)$ describing the asymptotic behavior of the model with respect to input size or hyperparameters~\cite{tunze2020sparsely,lin2022survey}.

Table~\ref{tab:complex} first shows $\mathcal{O}(\cdot)$ for the commonly used structures (layers) (i.e., multi-head self-attention (MHSA), Conv1d and Conv2d) for ECG recovery~\cite{tunze2020sparsely,lin2022survey}. In theory, Conv1d used by MMECG and RadarNet has a low time complexity with 1-D input and kernel size, while Conv2d in radarODE and RSSRNet significantly increases the complexity because the model needs to deal with extra information from the frequency domain. However, the comparison of $\mathcal{O}(\cdot)$ between different models can be trivial if multiple structures are used.

In Table~\ref{tab:complex}, FLOPs are also provided to directly show the calculations required for one forward propagation. The results indicate that radarODE and RSSRNet perform more calculations compared with others, because radarODE and RSSRNet are based on the spectrogram inputs, while MMECG and RadarNet are based on 1D radar signals inputs. In addition, RSSRNet adopts single-channel input while radarODE owns a large backbone to process $50$ channels as input, further increasing the time complexity.

\subsubsection{Space Complexity}
Space Complexity refers to the memory required to store weights and biases~\cite{guan2023achelous,tunze2020sparsely}, and can be easily revealed by counting the number of parameters in the model as shown in Table~\ref{tab:complex}. Similar to time complexity, RadarNet and MMECG have fewer parameters than RSSRNet and radarODE because of the 1-D input. The large parameter space of radarODE mainly comes from the powerful backbone with the ability to process multi-channel spectrogram inputs.

\subsubsection{Training Time per Epoch}
According to the complexity analysis above, the proposed radarODE has larger FLOPs and should spend more time for each epoch. However, the differences in training time per epoch are smaller than other metrics, because the training sets of other frameworks are formed based on arbitrary radar/ECG segments with a step length of $0.15$s, while the radarODE is based on the segment in each single cardiac cycle. Therefore, the number of segments in the training set of other frameworks is around $48$k and is much larger than that of radarODE with $19$k training pairs. The discrepancy in the scale of training dataset indicates that the proposed framework based on single cardiac cycles could alleviate the introduction of redundant information, because the segments based on step length may cause homogeneous training samples without any contribution to enhance the diversity of the dataset, but too many similar samples may increase the risk of overfitting and degrade the generalization ability of the deep learning model~\cite{schneider2024anchor}. 

\subsection{Discussions and Future work}
The proposed radarODE framework has shown outstanding performance compared with the previous work to generate faithful ECG signals under noisy scenarios, while the potential limitation will be discussed in this part to encourage future improvements in radar-based ECG recovery for real-life situations and applications.

\subsubsection{Long-range ECG Monitoring}
The direct impact brought by long-range monitoring is to reduce the SNR of the received radar signal according to the link budget analysis~\cite{liu2024diversity}. In addition, the cardiac location requires to be pre-identified to perform the accurate beamforming, because the current dataset assumes the majority of range-bins contain cardiac-related signals~\cite{chen2022contactless}.

\subsubsection{Complex Monitoring Scenarios}
Various new noises might be introduced and need to be eliminated, such as radar self-vibration introduced from car vibrations or hand-held radar~\cite{da2019theoretical}, mutual-radar interference for the future smart home with multiple electromagnetic devices~\cite{yang2024isense} and signal attenuation caused by human tissues for the monitoring of people with random body orientations~\cite{liu2024diversity}.

\subsubsection{Evaluation on the Dataset for Patients}
An important future application of radar-based ECG recovery is for clinical monitoring and diagnosis, while the ECG waveform for patients (e.g., arrhythmia) might be quite different, requiring massive new data for training. Some recent research has shown the feasibility of recovering abnormal ECG from radar signal~\cite{zhao2024airecg}, but more studies are required to investigate transfer learning or data augmentation techniques due to the scarcity of patient data. In addition, it is hard to preserve the noise-robustness for the ECG monitoring of patients, because the ODE model in this work is not designed for abnormal ECG patterns.

\section{Conclusions}\label{sec:conclusions}
Radar-based ECG reconstruction is highly reliant on purely data-driven approaches and lacks theoretical support regarding the transformation between mechanical activities measured by radar and electrical activities described as ECG. This research aims to bridge the gap to realize the transformation from the mechanical domain to the electrical domain by proposing the signal model with fine-grained features considered and further designing a deep learning framework radarODE with morphological prior embedding as ODEs. The radarODE framework is validated on the public dataset containing $4.5$ hours of radar measurements with corresponding ablation study and comparisons with the benchmark. The experimental result shows that radarODE could achieve a better MDR, morphological accuracy and peak accuracy than the benchmark, proving the rationality of the proposed signal model and the effectiveness of the radarODE under various physical statuses and random body movement. In the future, the complexity of the deep learning model needs to be reduced by squeezing the input size, and several data augmentation techniques might be applied to alleviate the data shortages, especially for patients.

\begin{IEEEbiography}[{\includegraphics[width=1in,height=1.25in,clip,keepaspectratio]{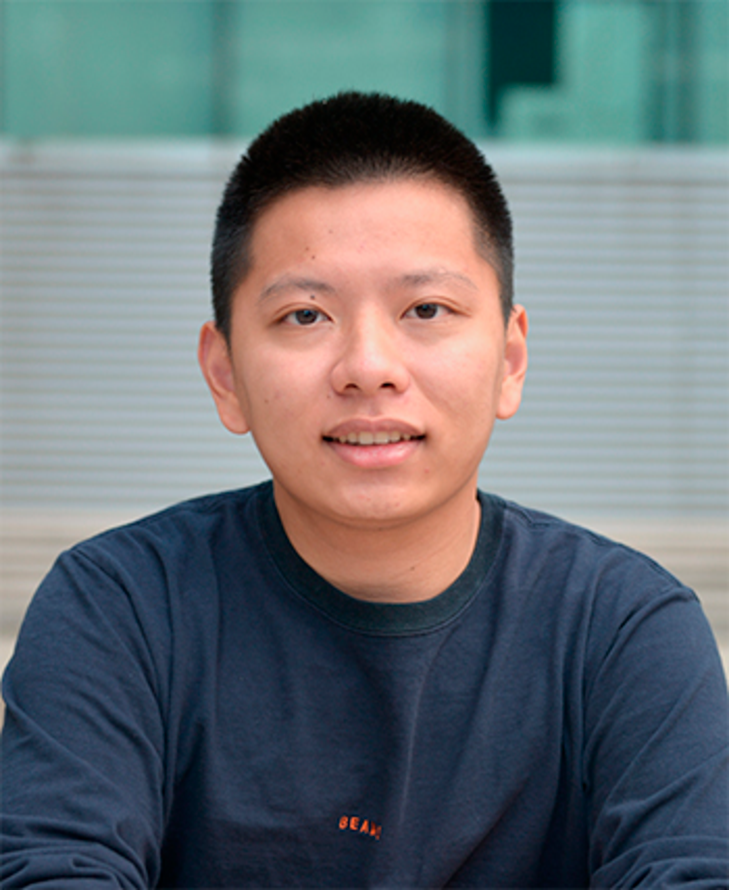}}]{Yuanyuan Zhang}
received the B.Eng. degree in electrical and electronic engineering from the University of Liverpool, UK, in 2020. He received M.S. degree in control system from the Imperial College London, UK, in 2021. He is currently pursuing the Ph.D. degree at the University of Liverpool, UK. His current research interests include contactless vital sign monitoring, wireless sensing and sparse signal processing. 
\end{IEEEbiography}

\begin{IEEEbiography}[{\includegraphics[width=1in,height=1.25in,clip,keepaspectratio]{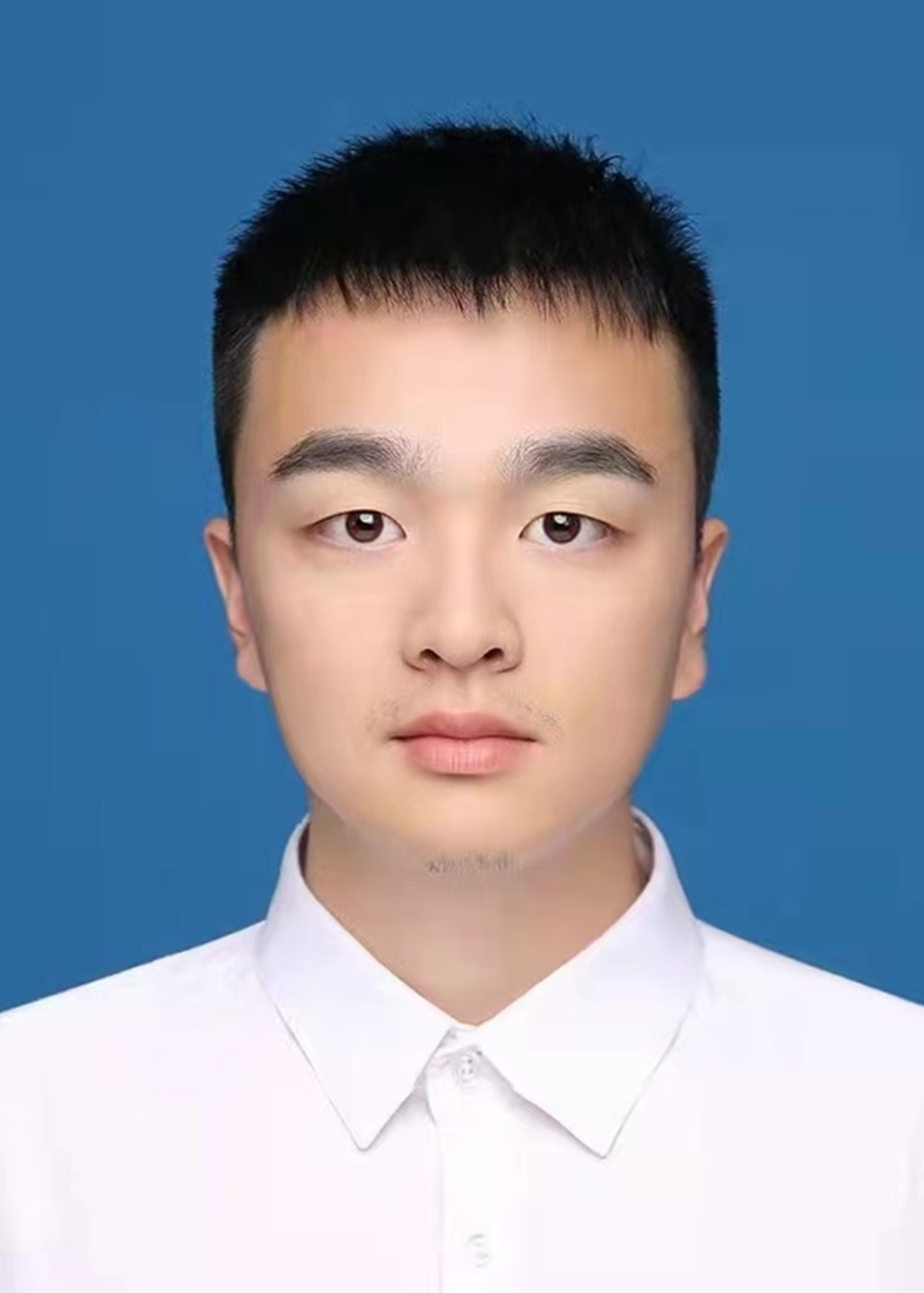}}]{Runwei Guan} is currently a joint Ph.D. student of University of Liverpool, Xi'an Jiaotong-Liverpool University and Institute of Deep Perception Technology, Jiangsu Industrial Technology Research Institute. His research interests include multi-modal perception, lightweight neural network and statistical machine learning. He has published more than 10 papers of SCI/CCF/CAA/EI. He serves as a peer reviewer of IEEE TNNLS, TITS, TCSVT, ITSC, EAAI, NeuroCom., MTAP, etc.\end{IEEEbiography}

\begin{IEEEbiography}[{\includegraphics[width=1in,height=1.25in,clip,keepaspectratio]{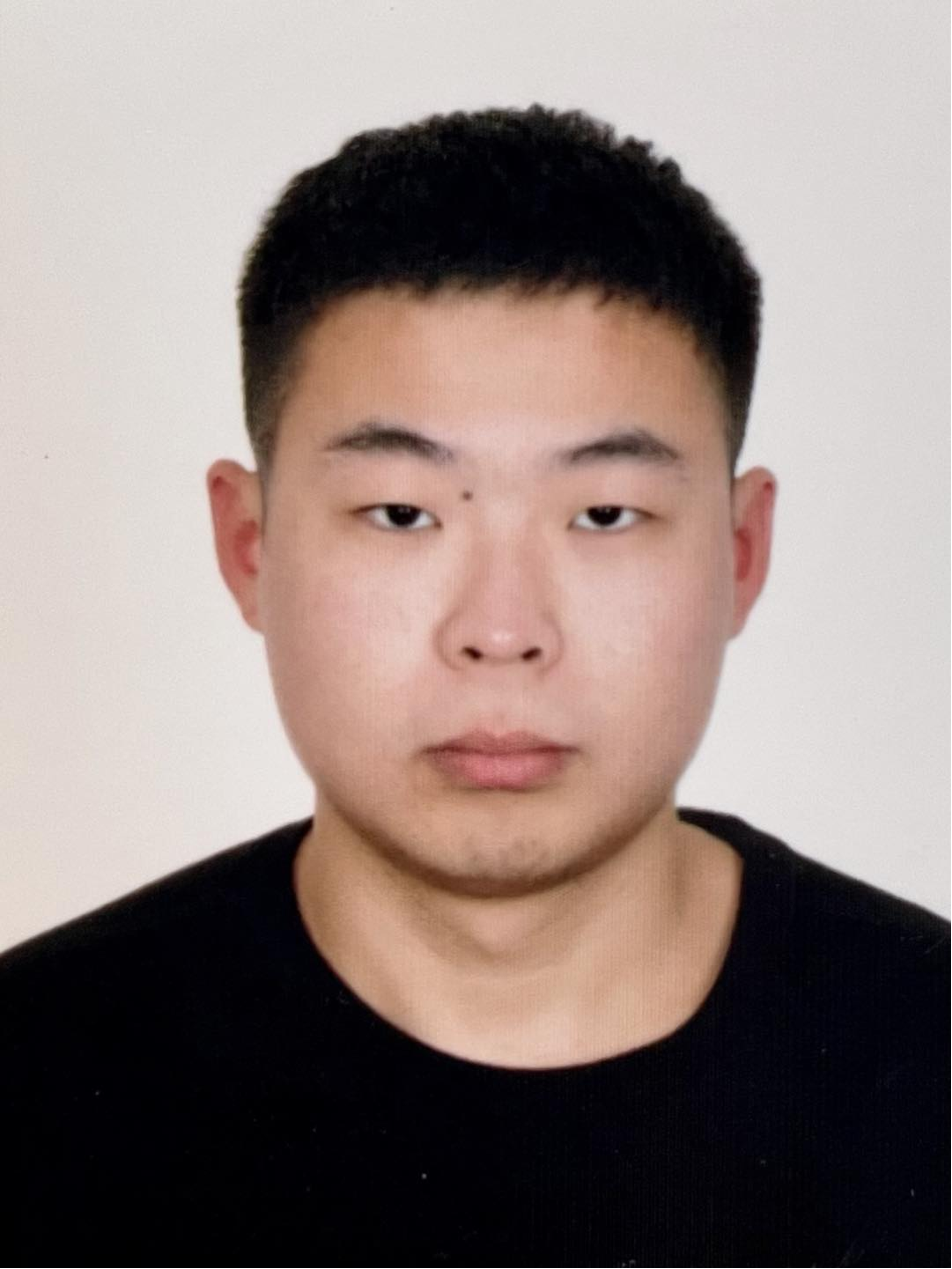}}]{Lingxiao Li}
received the B.S. degree in Computer Science from the University of Liverpool, UK, in 2020. He received M.S. degree in Computer Science from the Columbia University, NY, USA, in 2022. He is currently a research assistant at the Multimedia Lab (MMLab), Department of Information Engineering, the Chinese University of Hong Kong. His current research interests include computer vision and machine learning. 
\end{IEEEbiography}

\begin{IEEEbiography}[{\includegraphics[width=1in,height=1.25in,clip,keepaspectratio]{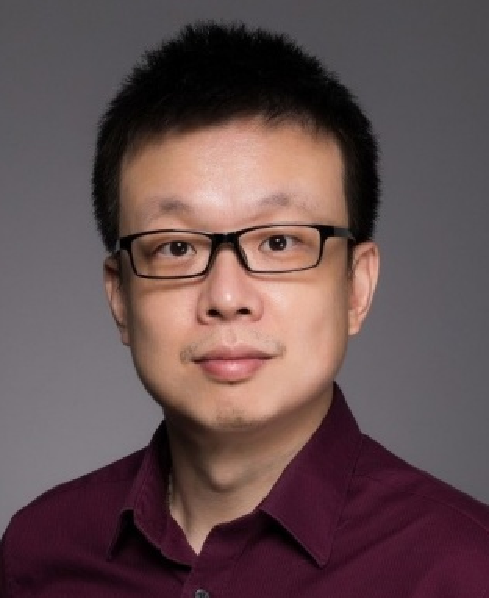}}]{Rui Yang} received the B.Eng. degree in Computer Engineering and the Ph.D. degree in Electrical and Computer Engineering from National University of Singapore in 2008 and 2013 respectively. 

He is currently an Associate Professor in the School of Advanced Technology, Xi’an Jiaotong-Liverpool University, Suzhou, China, and an Honorary Lecturer in the Department of Computer Science, University of Liverpool, Liverpool, United Kingdom. His research interests include machine learning based data analysis and applications. Dr. Yang is currently serving as an Associate Editor for IEEE Transactions on Instrumentation and Measurement, Neurocomputing, and Cognitive Computation.
\end{IEEEbiography}
\vfill
\begin{IEEEbiography}[{\includegraphics[width=1in,height=1.25in,clip,keepaspectratio]{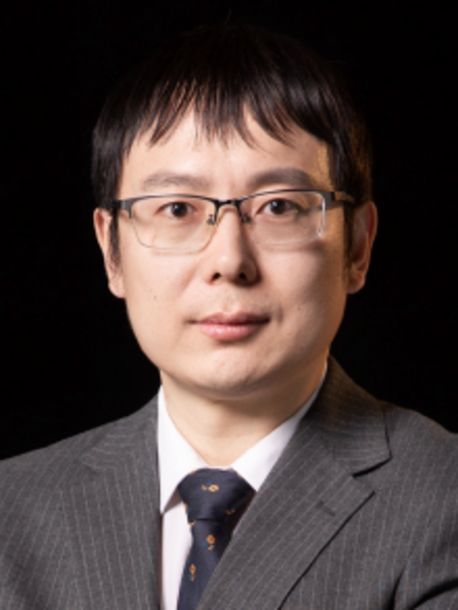}}]{Yutao Yue} (Senior Member, IEEE) is an associate professor at the Artificial Intelligence Thrust and Intelligent Transportation Thrust of Hong Kong University of Science and Technology (Guangzhou). He received his Bachelor’s degree from the University of Science and Technology of China, and Master and PhD degree from Purdue University. He has a dual background in academia and industry, as the team leader of Guangdong Province Introduced Innovation Scientific Research Team, senior scientist of Kuang-Chi Group, and the founder of the Institute of Deep Perception Technology of JITRI. His research interests include multimodal perception fusion, machine consciousness, artificial general intelligence, causal emergence, etc. He has been engaged in scientific research and technology industrialization for over 20 years. He has co-invented 354 granted Chinese patents, 18 USA patents, and 7 EU patents. He has led 6 major research projects with a total funding of nearly 130 million RMB. He has published over 60 papers, advised 13 postdoc research fellows, and received multiple awards including Wu Wenjun Artificial Intelligence Science and Technology Award.
\end{IEEEbiography}
    
\begin{IEEEbiography}[{\includegraphics[width=1in,height=1.25in,clip,keepaspectratio]{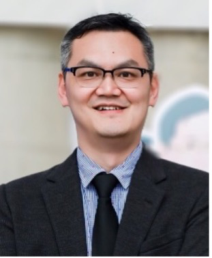}}]{Eng~Gee~Lim}
(M'98-SM'12) received the BEng(Hons) and PhD degrees in Electrical and Electronic Engineering from UK in 1998 and 2002 respectively. Prof. Lim worked for Andrew Ltd, a leading communications systems company in the United Kingdom from 2002 to 2007. Since August 2007, Prof. Lim has been at Xian Jiaotong-Liverpool University, where he was formally the head of EEE department and University Dean of Research and Graduate studies. Now, he is School Dean of Advanced Technology, director of AI university research centre and also professor in department of Communications and Networking. He has published over 200 refereed international journal and conference papers. His research interests are Artificial Intelligence, robotics, AI+ Health care, Future Education, Management in Higher Education, International Standard (ISO/IEC) in Robotics, antennas, RF/microwave engineering, EM measurements/simulations, energy harvesting, power/energy transfer, smart-grid communication; wireless communication networks for smart and green cities. He is a charter engineer and Fellow of both IET and Engineers Australia. In addition, he is also a senior member of IEEE and Senior Fellow of HEA.
\end{IEEEbiography}
\vfill
\end{document}